\documentclass[10pt,journal,compsoc]{IEEEtran}

\ifCLASSOPTIONcompsoc
  \usepackage[nocompress]{cite}
\else
  \usepackage{cite}
\fi
\usepackage[font=sf]{caption}
\usepackage[utf8]{inputenc}
\usepackage{acronym}
\usepackage[british]{babel}
\usepackage{color}
\usepackage{graphicx}
\usepackage{subfig}
\usepackage{amsmath}
\usepackage{amsthm}
\usepackage{amssymb}
\usepackage{algpseudocode}
\usepackage{algorithm}
\usepackage{listings}
\usepackage{array}
\usepackage{balance}
\usepackage{enumitem}
\usepackage{nth}
\usepackage{url}

\usepackage{tikz}
\usepackage{pgfplots}
\usepackage{pgfplotstable}
\usepgfplotslibrary{groupplots}

\DeclareMathOperator{\ADD}{\textsc{Add}}
\DeclareMathOperator{\LOOKUP}{\textsc{Lookup}}
\DeclareMathOperator{\ENCODE}{\textsc{Encode}}
\DeclareMathOperator{\DECODE}{\textsc{Decode}}

\theoremstyle{definition}
\newtheorem{example}{Example}[section]
\newtheorem{lemma}{Lemma}[section]

\acrodef{FP}{False Positive}
\acrodef{FN}{False Negative}

\newcommand{\todo}[1]{}
\renewcommand{\todo}[1]{{\color{red} {#1}}}

\begin{document}

\title{Multiple Set Matching and Pre-Filtering with Bloom Multifilters}
\author{Francesco Concas, Pengfei Xu, Mohammad A. Hoque, Jiaheng Lu, and Sasu Tarkoma}

\IEEEtitleabstractindextext{
\begin{abstract}
  Bloom filter is a space-efficient probabilistic data structure for checking elements' membership in a set. Given multiple sets, however, a standard Bloom filter is not sufficient when looking for the items to which an element or a set of input elements belong to. In this article, we solve multiple set matching problem by proposing two efficient Bloom Multifilters called Bloom Matrix and Bloom Vector. Both of them are space efficient and answer queries with a set of identifiers for multiple set matching problems.  We show that the space efficiency can be optimized further according to the distribution of labels among multiple sets: Uniform and Zipf.  While both of them are space efficient, Bloom Vector can efficiently exploit Zipf distribution of data for further space reduction. Our results also highlight that basic $\ADD$ and $\LOOKUP$ operations on Bloom Matrix are faster than on Bloom Vector. However, Bloom Matrix does not meet the theoretical false positive rate of less than $10^{-2}$ for $\LOOKUP$ operations if the represented data or the labels are not uniformly distributed among the multiple sets. Consequently, we introduce \textit{Bloom Test} which uses Bloom Matrix as the pre-filter structure to determine which structure is suitable for improved performance with an arbitrary input dataset. 
  

\end{abstract}
}

\maketitle

\setlength{\textfloatsep}{0.1cm}
\setlength{\floatsep}{0.1cm}

\renewcommand{\baselinestretch}{0.98}
\section{Introduction} \label{introduction}

Modern popular Internet services, including Google search, Yahoo directory, and web-based storage services, rely on efficient data matching \cite{BrP12}. They have custom techniques for providing scalable, fault-tolerant and low-cost services \cite{PBC00, Rab89, SKH95}. Fast matching of arbitrary identifiers to specific values is a fundamental requirement of these applications in which data objects are organized using unique local or global identifiers, usually called labels \cite{KaT06}.  In a typical usage scenario, each label maps to a set of values \cite{KaT13}. For example, in order to provide low latency for accessing data, such data is cached across different regions; given a set of contents, the application needs to find to which proxy servers they are mapped to.

A popular probabilistic data structure called Bloom Filter serves a similar purpose, namely to answer whether a label belongs to a particular set. The standard Bloom Filter represents a set of labels, using a number $n$ of bits. Although it is a very space efficient structure, it can only answer whether a label $l$ belongs to a set $e$, with a probability of false positives. There are many extensions of the Bloom Filter, such as space-code~\cite{KXW06}, spectral~\cite{CoM03}, and Shift Bloom Filter~\cite{YLS16}. These structures represent a multiset, where a label can exist multiple times. Therefore, they can answer multiplicity queries, for example, how many times a label exists in a multiset. 

 Let $L = \{l_1, ..., l_{|L|}\}$ be a set of labels and $E = \{e_1, ..., e_{N}\}$ be a set of items. We are interested in representing the function $f : L \rightarrow \mathcal{P}(E)$, where $\mathcal{P}(E)$ is the power set of $E$.  Nevertheless, all the mentioned extensions are not able to locate \emph{multiple sets} when finding a single or multiple labels. In this study, we present two data structures, namely Bloom Matrix and Bloom Vector, that can represent such relations for \emph{multiple sets} where each set contains a number of unique elements.  While both structures use multiple Bloom Filters to represent multiple sets, Bloom Matrix contains the Bloom Filters of equal size, whereas the Bloom Filters in Bloom Vector can be of different sizes.

These new data structures associate multiple sets and  support queries with single or  multiple labels through an inverted indexing solution using multiple Bloom Filters, while providing a reasonable accuracy. The most efficient $\LOOKUP(l)$ operation on them returns a set or list of items, i.e., $S\in \mathcal{P}(E)$, rather than a simple true or false answer.  

\subsection{Example applications}
The usage of Bloom Filters in networking is widespread. They are suitable to summarize the contents of a P2P network for supporting collaborative operations~\cite{CPM03}. They also have been applied to enhance probabilistic algorithms for locating resources~\cite{ReV03}, for taking routing decisions~\cite{RhK02}, and for traffic monitoring~\cite{TBJ15}. 
In this subsection, we discuss some applications in which Bloom Multifilters could be useful: load balancing,  web caching, and document search.

\vspace{1mm}
\noindent\textbf{Load balancing.} Load balancing is used to optimally distribute workloads among multiple computing resources, in order to maximize the performance of the whole system \cite{Rab89, Cyb89, BaW89, SKH95}. 

If a cluster of servers provides multiple services, with those services distributed across different servers, the services can be represented as the labels $L$ and the servers can be represented as the items $E$, and one of the proposed BMFs can be used to map the services to the servers that provide them.

\vspace{1mm}
\noindent\textbf{Web caching.} Much of the modern Internet is composed of proxy servers. This allows the workload to be split, also providing redundancy and an extra level of security due to the provider being hidden by the proxy servers. Whenever a client needs to access a service from the provider, the request from the client is handled by one of the proxy caches. 

 In a proxy cache, Internet contents tend to be small compared to the number of requests. If we represent the contents as labels $L$ and the proxy servers spread across as the set of items $E$, we could apply our Bloom Multifilters to the proxy caches for a service.


\vspace{1mm}
\noindent\textbf{Document Search.} Document search with keywords is another important use case. In a search engine, such as Solr\footnote{\url{http://www.solrtutorial.com/basic-solr-concepts.html}}, the documents go through multiple transformations while being added, and a series of tokens are generated from different fields of the documents. These tokens are added to the index. In this way, Solr achieves faster document retrieval. However, indexing different fields increases the size of the index and slows down the search.

The proposed Bloom Multifilter structures can be applied as an alternate solution to represent large text corpora. For example, the Wikipedia corpus can be represented as the set of all pages or items, i.e., $E$. The unique words in a page can be the set of labels, $L$, which can be represented as a Bloom Filter for a document. These Bloom Filters could also be encoded as Solr indices rather than the actual words. A search query would returns a list of page identifiers.

\subsection{Contributions}
In this article, we present two new Bloom Multifilters (BMFs)  for fast space-efficient matching of arbitrary identifiers to sets, at the cost of introducing a false positive rate, similarly to Bloom Filters. Our contributions are the following:

\begin{itemize}[leftmargin=*]
    \item We introduce two new Bloom Multifilters, namely Bloom Matrix and Bloom Vector, to solve the labels-to-sets matching problem. These data structures are inspired by standard Bloom Filter and some of its extensions (see Section~\ref{relatedwork}), in particular, the Bloomier filter \cite{CKR04} and the Bloom Multifilter \cite{XuR16}. Bloomier filter encodes only one set without false positives, whereas our Bloom Matrix and Vector encode multiple sets and can have false positives. Compared to the other Bloom Multifilter \cite{XuR16}, our Multifilters return the set of items or their identifiers instead of a simple true or false answer for a $\LOOKUP$ operation.

\item We present theoretical analysis and demonstrate that they adhere basic Bloom Filter operations. We evaluate the performance of these structures with different configuration parameters and the distribution of the labels in the presence of synthetic and 20 Newsgroups corpus test dataset \cite{Car07}. Our results also highlight that basic $\ADD$ and $\LOOKUP$ operations on Bloom Matrix are faster than on Bloom Vector. While both of them are space efficient, Bloom Vector can efficiently exploit Zipf distribution of data for further space reduction. 
  
   \item We also evaluate their performance according to the desired false positive rate with a dataset of unknown distribution of labels. We demonstrate that Bloom Matrix can be used to determine whether such dataset follows Uniform distribution or not. We introduce \emph{Bloom Test} to determine the appropriate Multifilter according to a desired false positive rate lower than $10^{-2}$. In other words, Bloom Matrix can be used as a pre-filter structure to model the input data with multiple sets.

\end{itemize}

We organize the rest of the article as follows. In Section~\ref{preliminaries}, we introduce the readers to standard Bloom Filter and its properties. In Section~\ref{problem}, we describe the problem. Section~\ref{bloom_matrix} and Section~\ref{bloom_vector} present the definitions and theoretical analysis of the Bloom Matrix and Vector. In Section~\ref{experimental_results}, we investigate the performance of our Bloom Multifilters and introduce Bloom Test in Section~\ref{bloomtest}. We outline the related works in Section~\ref{relatedwork} before concluding the paper. 

\section{Preliminaries} \label{preliminaries}

\begin{table}[!h]
    \centering
    \footnotesize
    \begin{tabular}{| l |p{6.5cm} |}
        \hline
        Notations & Implications\\\hline
        $l, L$  & a label and a set of labels \\\hline
        $e, E$  & an item and set of items \\\hline
        $h, H$  & a hash function and a set of hash functions \\\hline
        $M$  & the range of hash neighborhoods \\\hline
        $N$  & the total number of items \\\hline
        $m$  & the number of bits in one Bloom Filter, calculated from Eq. \ref{mbf} \\\hline
        $k$  & the number of hash functions in one Bloom Filter, calculated from Eq. \ref{kbf} \\\hline
        $n$  & the number of items inserted to the structure specified in the context \\

        \hline
    \end{tabular}
    \caption{Notations used for Bloom Multifilters.}
    \label{tab:notation}
\end{table}

Bloom Filter \cite{Blo70} is a probabilistic data structure which represents a set so that the set will occupy much less memory space than it normally would when represented with conventional methods. This comes at the cost of introducing a \ac{FP} rate. \acp{FN}, on the other hand, are not allowed. In this section, we outline the Bloom Filter and other variations of Bloom Filter. Table~\ref{tab:notation} summarizes the notations used to describe Bloom Filter and Bloom Multifilters in this article. 

\subsection{Bloom Filter}



\vspace{1mm}
\noindent\textbf{Definition.} Let $U$ be a set of all the items that we can possibly store, and let $E \subseteq U$ be the set that we wish to represent. We are interested in encoding the function $f : U \rightarrow \{0, 1\}$ defined as:
\[
    f(x) =
    \begin{cases}
        1 & \text{if } x \in E \\
        0 & \text{otherwise}
    \end{cases}
\]

The Bloom Filter is defined as a pair $(B, H)$, where $B$ is a bitset of size $m$ and $H = \{h_1(x), ..., h_k(x)\}$ is a set of hash functions, each having image $[0, m-1]$. There are two operations on Bloom Filter: $\ADD(x)$ and $\LOOKUP(x)$. A remove or delete operation on Bloom Filter would introduce a chance of FN.

The set of distinct values returned by all the hash functions for an input label $x$ is called its \textit{hash neighborhood}; we define it as $H(x)$ with the abuse of notation.

\vspace{1mm}
\noindent\textbf{Add.} When a Bloom Filter is created, all the bits in bitset $B$ are initialized to 0. Whenever we add a label or element $x$, we set to 1 each $B[i]$ for each $i \in H(x)$.


\vspace{1mm}
\noindent\textbf{Lookup.} To test the membership of an element $x$, we have to check whether all of the bits $B[i]$ for each $i \in H(x)$ are set to 1. If it is true, then the element is probably in the set; otherwise, it is definitely not in the set.


\vspace{1mm}
\noindent\textbf{False positive rate.} FPs occur whenever we look for an element $x$ which is not in the set, and the $\LOOKUP$ function returns true. Such function returns true whenever all the bits are having as indices the neighborhood of $x$ is set to 1. This implies that the more bits are set to 1, the higher is the FP rate. The number of hash functions also influences the FP rate. A higher number of hash functions decreases the chance of collisions between two different elements. Therefore, the choice of an optimal number of hash functions is a compromise.

Bose et al. \cite{BGK08} have shown that the probability $p$ of false positives in a Bloom Filter of size $m$, $k$ hash functions, and have added $n$ elements is:
\begin{align}\label{fpratebf}
    p = \Theta\left(\left[1 - \left(1 - \frac{1}{m}\right)^{kn}\right]^k\right) = \Theta\left(\left(1-e^{-kn/m}\right)^k\right)
\end{align}

If we know a priori the number of elements that we are going to insert in a Bloom Filter, we can choose its parameters so that the Bloom Filter will have a probability of false positives around a certain value $p$. We derive from Equation~\ref{fpratebf}:
\begin{subequations}
    \begin{equation}\label{mbf}
        m = - n\frac{\ln p}{\ln^2{2}}
    \end{equation}
    \begin{equation}\label{kbf}
        k = \ln 2 \cdot \frac{m}{n} = - \log_2 p
    \end{equation}
\end{subequations}

\section{Problem Definition} \label{problem}


Standard Bloom Filter and it's extensions can encode only one set and can answer whether a label belongs to a set or not. In this article, we extend standard Bloom Filter not only to encode multiple sets and to efficiently check the membership of an element in all the sets but also to answer to which of the sets such element belongs to, i.e. a set of identifiers. We discuss the related works in Section~\ref{relatedwork}.

\subsection{Definition}


Let $L = \{l_1, ..., l_{|L|}\}$ be a set of labels and $E = \{e_1, ..., e_{N}\}$ be a set of items. We are interested in representing the function $f : L \rightarrow \mathcal{P}(E)$, where $\mathcal{P}(E)$ is the power set of $E$.

The most straightforward approach is to use a Bloom Filter to store the labels associated with each item. We also show another approach that has both advantages and disadvantages compared with the former. The idea of the latter approach is to represent the function $f$ similarly to a Bloom Filter. However, instead of using single bits to encode the elements, we use bitsets in which we store binary representations of the element $s \in \mathcal{P}(E)$ to which the labels map to. We obtain these representations with two functions; $\ENCODE$ and $\DECODE$.

\subsection{Encode and Decode}

Let $\Pi$ be an ordering on $E$. We introduce  $\ENCODE(\Pi,S)$, which returns a binary representation $V = \{ v_1, ..., v_N \}$ of $S$ given an ordering $\Pi$ of $E$ and a set of items $S \in \mathcal{P}(E)$, such that:
\[
    v_i =
    \begin{cases}
        1 & \text{if } \Pi(i) \in S \\
        0 & \text{otherwise}.
    \end{cases}
\]
We also introduce $\DECODE(\Pi,V)$, which returns $S$ given an ordering $\Pi$ of $E$ and a binary representation $V$ of a set of items $S \in \mathcal{P}(E)$.

$\ENCODE$ and $\DECODE$ essentially associate each element to a bit in a binary representation, as we can see in the following Example~\ref{ex:encdec}:

\begin{example} \label{ex:encdec}
    Let $E = \{e_1, e_2, e_3\}$ and $\Pi$ being an ordering on $E$, which in this case orders the element as written in the definition of $E$. We have:
    \[ \ENCODE(\Pi, \{e_1, e_3\}) = \{1,0,1\}\]
    \[\DECODE(\Pi, 011) = \{e_2, e_3\} \]
\end{example}



\vspace{1mm}
\noindent\textbf{Analysis of Encode and Decode.} The $\ENCODE(\Pi, S)$ function needs to initialize a bitset $V$ to 0 and to set some bits to 1, in order to return the encoded value of $S$. The space complexity is, therefore, $\Theta(N)$. For each element in $S$, we need to set $V[i] \gets 1$, where $i$ is the index of $S$ in the ordering $\Pi$. Since we are using $\Pi$, the time complexity is $O(|S|)$.

The $\DECODE(\Pi, V)$ function needs to create a set from the bitset $V$, which can be at most $N$. Therefore it takes $O(N)$ space. For each bit $v_i$ set to 1 in $V$, the function fetches the element at position $i$ in the array and adds it to the set. If we need to return an ordered set, it takes $O(|V| \log |V|)$ time; otherwise it takes $O(|V|)$ time. In the rest of this thesis, we assume that we do not need to return an ordered set.


\section{Bloom Matrix} \label{bloom_matrix}

In this section, we introduce \textit{Bloom Matrix} as our first effort to solve the membership checking problem with multiple sets. Precisely, it consists of multiple columns of bitsets, in which each column represents an item, and values of its bits are determined by associated labels, e.g., the set of unique words (as labels) in a document (as an item). In other words, unlike  Bloom  Filter,  each  bit  in  the  Filter  is  replaced  by another  bitset  of  a  fixed  length,  hence  the  name  is  Bloom Matrix.


\subsection{Definition}

We define a Bloom matrix as a triplet $(\mathbf{G}, \Pi, H)$, where $\mathbf{G}$ is a binary matrix of size $m \times N$, $\Pi$ represent an ordering on the set $E$ as previously defined, and $H = \{h_1(x), ..., h_k(x)\}$ is a set of hash functions, each having image $[0,m-1]$. If using MurmurHash, we can replace $H$ with a number of hash functions $k$, and use as seeds for MurmurHash the range $[1,k]$. In the rest of the thesis we replace $H$ with $k$.

\subsection{Operations}


\vspace{1mm}
\noindent\textbf{Add.} $\mathbf{G}$ is initialized with all its bits set to 0. In order to add a label $l$ to a Bloom Matrix $F$, we add the value returned by $\ENCODE(\Pi, f(l))$ to the rows in the bit matrix $\mathbf{G}$ having the indices equal to the hash neighborhood of $l$, using the bitwise OR operator. The add operation can be formally defined as:
\begin{align*}
    F.\ADD(l) &:= \mathbf{G}[h_i(l),\_] \gets \mathbf{G}[h_i(l),\_] \vee \ENCODE(\Pi, f(l)) \\
    & \text{ for } 1 \leq i \leq k
\end{align*}

 The steps are illustrated in Alg.~\ref{algo:bmadd}. Suppose that we want to add some label $l$ to the Bloom Matrix. We obtain a bitset, $V$, from $\ENCODE(\Pi, f(l))$ in which the bits set to 1 are at the positions $V = \{v_1, ..., v_{|V|}\}$, which represent columns of the Bloom matrix. We next obtain a set of indices $H = \{h_1(l), ..., h_k(l)\}$ from the hash functions, which represent rows of the Bloom Matrix. Therefore, the $\ADD(l)$ function is going to set to 1 all the bits whose indices are given by the Cartesian product $V \times H$.

\vspace{1mm}
\noindent\textbf{Lookup.} In order to find out which subset of $\mathcal{P}(E)$ is labelled with $l$, we use the $\DECODE$ function on the bitset resulting from the bitwise AND operation on the rows in $\mathbf{G}$ having the indices equal to the hash neighborhood of $l$:
\[ F.\LOOKUP(l) := \DECODE \left(\Pi, \bigwedge_{1 \leq i \leq k} \mathbf{G}[h_i(l),\_] \right) \]

\vspace{1mm}
\noindent\textbf{Multiple labels lookup.} The lookup operation is illustrated in Alg.~\ref{algo:bmlookup}. Similarly, we can lookup for multiple labels by computing the hash neighborhood of all the labels, the rest of the lookup algorithm is identical to the algorithm for looking up a single label.


\begin{figure}
    \centering
    \includegraphics[width=0.8\linewidth]{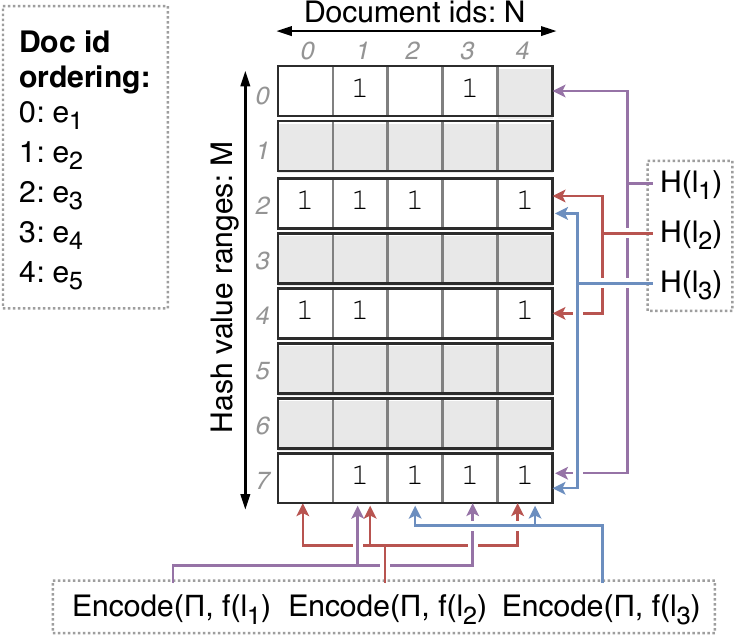}
    \caption{Example of a Bloom matrix $\mathbf{G}[m,N]$ obtained from Example~\ref{BMexample}.}
    \label{fig:BM}
\end{figure}

\begin{algorithm}[h!]
    \caption{Bloom Matrix: Add Operation}
    \begin{algorithmic}[1]
        \Procedure{Add}{$l$}
            \State $V \leftarrow \Call{Encode}{\Pi, f(l)}$
            \State $H \leftarrow \Call{GetNeighbourhood}{l, k, m}$
            \For {$i \leftarrow H$}
                \State $\mathbf{G}[i,\_] \gets \mathbf{G}[i,\_] \vee V$
            \EndFor
        \EndProcedure
    \end{algorithmic}
    \label{algo:bmadd}
\end{algorithm}

\begin{algorithm}[h!]
    \caption{Bloom Matrix Lookup}
    \begin{algorithmic}[1]
        \Procedure{Lookup}{$l$}
            \State $H \leftarrow \Call{GetNeighbourhood}{l, k, m}$
            \State Let $V$ be an empty bitset
            \For {$i \leftarrow H$}
                \State $V \leftarrow V \wedge \mathbf{G}[i,\_]$
            \EndFor
            \State \Return $\Call{Decode}{\Pi, V}$
        \EndProcedure
    \end{algorithmic}
        \label{algo:bmlookup}
\end{algorithm}

\begin{example}\label{BMexample}
    Given 3 labels $L = \{l_1, l_2, l_3\}$, 5 items $E = \{e_1, \cdots, e_5\}$, an ordering $\Pi$ that sort items in input order, and two hash functions $H = \{h_1(x), h_2(x)\}$ returns $[0,7]$ such that $H(l_1) = \{0,7\}$, $H(l_2) = \{2,4\}$ and $H(l_3) = \{2,7\}$. The labels can be assumed as the words and items can be assumed as the documents.
    
    We now represent $f(l_1) = \{e_2, e_4\}$, $f(l_2) = \{e_1, e_2, e_5\}$ and $f(l_3) = \{e_3, e_5\}$ with a Bloom Matrix. As showing in Figure \ref{fig:BM}, when adding $l_1$, we determine target rows $\mathbf{G}[0,\_]$ and $\mathbf{G}[7,\_]$ according to $H(l_1) = \{0,7\}$, then run $\ENCODE$ function to find two columns $\mathbf{G}[\_,2]$ and $\mathbf{G}[\_,4]$ since $\ENCODE$ returns $\{0,1,0,1,0,0\}$. We finally set four bits to $1$: $\mathbf{G}[0,2]$, $\mathbf{G}[7,2]$, $\mathbf{G}[0,4]$ and $\mathbf{G}[7,4]$. $l_2$ and $l_3$ can be added by the same manner.
    
    When preforming $\LOOKUP(l_1)$, we do hash operation $H(l_1)=\{0,7\}$ and then AND two rows: $\mathbf{G}[0,\_]$ and $\mathbf{G}[7,\_]$. This operation returns $\{0,1,0,1,0,0\}$. Therefore, $\DECODE$ returns $\{e_2,e_4\}$. Noteworthy, $\LOOKUP(l_3)$ perform AND operation on $\mathbf{G}[2,\_]$ and $\mathbf{G}[7,\_]$ which outputs $\{e_2, e_3, e_5\}$, with a \ac{FP} $e_2$.
\end{example}

\vspace{1mm}
\noindent\textbf{Update.} Bloom Matrix is a fixed structure for the sets. It is not possible to update the Matrix with the labels of a new item and it  requires reconstruction of the Matrix. However, it possible to add a label to an existing item at a cost of an increasing FP rate.

\subsection{False Positive Rate}

\begin{lemma}
    Given a set of labels $L=\{l_1,\cdots,l_{|L|}\}$ where each $l$ is associated with some items $e$. The total false positive rate of a Bloom Matrix when performing multiple label lookup on L, $\LOOKUP(L)$ is 
    \begin{equation}\label{eq:BMFP_lemma}
        \overline{\text{FPR}} =\frac{\sum_{l\in L} \sum_{e\in E\setminus f(l)} (1-(1-\frac{1}{m})^{nk})^k}{|L|}
    \end{equation}
    where $k$ is the number of hash functions $m$ is the range of hash neighborhoods, $n$ is the number of items added to the Bloom Matrix, and $E$ is the universe of all items.
\end{lemma}
\begin{proof}
    Given $k$ hash functions, Bloom Matrix's $\LOOKUP$ operation performs AND operation on $k$ rows, which indexes determined by the value of hash neighborhoods. Therefore, a false positive in this case is when the bit at column $c$, i.e., bits of a document id in Figure \ref{fig:BM}, of all $k$ rows are set to $1$, where $c$ is a positive integer.

    When executing $\ADD$ operation for a label $l$, a hash function sets bits uniformly to one of $m$ rows for each item $e \in f(l)$. Therefore, for any single bit in Bloom Matrix, the probability that it is not set to $1$ by $k$ hash functions during one $\ADD$ operation is 
    \begin{equation}
        \label{eq:bm_fp_1}
        (1-\frac{1}{m})^k
    \end{equation}

    Assume that $\ADD$ operation is performed on $n = |f(l)|$ items, i.e., we have added a label $n$ items, then probability that the bit is still $0$ is
    \begin{equation}
        \label{eq:bm_fp_1a}
        (1-\frac{1}{m})^{nk}\nonumber
    \end{equation}

    In contrast, the probability that the bit is set to $1$ is
    \begin{equation}
        \label{eq:bm_fp_2}
        1-(1-\frac{1}{m})^{nk}
    \end{equation}

    Now assume that there is a false positive item $e_f$ assigned to a label $l$ for the $\LOOKUP$ operation. Then, there are $k$ bits at a single column, whose index corresponding to $e_f$, and $k$ rows, whose index determined by hash functions, are set to $1$. This happens with probability
    \begin{equation}
        \label{eq:bm_fp_3}
        (1-(1-\frac{1}{m})^{nk})^k
    \end{equation}

    In other words, when we perform $\LOOKUP$ operation for a label $l$, each returned items has a probability equals to Equation \ref{eq:bm_fp_3} for being a false positive. Since $e_f$ is the member of $e\in E\setminus f(l)$, then any member of $E\setminus f(l)$ has the probability of being false positive, i.e.,
\begin{equation}
 \sum_{e\in E\setminus f(l)} (1-(1-\frac{1}{m})^{nk})^k.
\end{equation}

Furthermore, for multiple label lookup in $L$, the overall false positive rate for looking up all labels $l \in L$ can be obtained by summing up the probabilities for all items not in any $f(l)$, and the average false positive rate is,
    \begin{equation}\label{eq:BMFP}
        \overline{\text{FPR}} =\frac{\sum_{l\in L} \sum_{e\in E\setminus f(l)} (1-(1-\frac{1}{m})^{nk})^k}{|L|}
    \end{equation}
\end{proof}

\subsection{Complexity}

\begin{lemma}
    Bloom Matrix has a space complexity $\Theta(mN)$, $\ADD$ time complexity $O(N \log |e|)$ and $\LOOKUP$ complexity $\Theta(kN)$, where $m$ is the size of hash neighborhood, $k$ is the number of hash functions, $N$ is the \textit{total} number of items and $|e|$ is the number of input items in an $\ADD$ operation.
\end{lemma}
\begin{proof}
    Space complexity: Bloom Matrix stores three components: (i) the bitset that has size $m \times N$, (ii) The total ordering $\Pi$, which has size $2N$, required by $\ENCODE$ and $\DECODE$ functions, and (iii) $k$ hash functions. The space cost is therefore $\Theta(mN) + 2\Theta(N) + \Theta(k) = \Theta(mN)$ when $k \ll mN$.
    
    Time complexity: an $\ADD(l, e)$ operation on the Bloom Matrix computes the neighborhood of $l$ and executes $\ENCODE(\Pi, e)$ to obtain a bit sequence $V$, then updates the matrix according to $V$. The $\ENCODE$ takes $\Theta(|e|)$ time, while the hash operations take $\Theta(k)$ time in total. Then, insertion in the matrix takes $\Theta(N)$ time because there are exactly $k$ bits in each column altered (with a constant operate time), while there are at most $N$ columns. The time of $\ADD$ is therefore $\Theta(k) + \Theta(|e|) + \Theta(k) = \Theta(|e|+k)$.

    A $\LOOKUP(l)$ operation computes the neighborhood of $l$ and executes $\DECODE(\Pi, V)$ on the $V$ obtained by bitwise AND operation on the $k$ rows, each has at most $N$ bits.
    The time cost is $\Theta(N) + \Theta(kN) = \Theta(kN)$.
\end{proof}



\subsection{Sparse Bloom Matrix}
\label{sec:sbm}

\begin{figure}
    \centering
    \includegraphics[width=.5\linewidth]{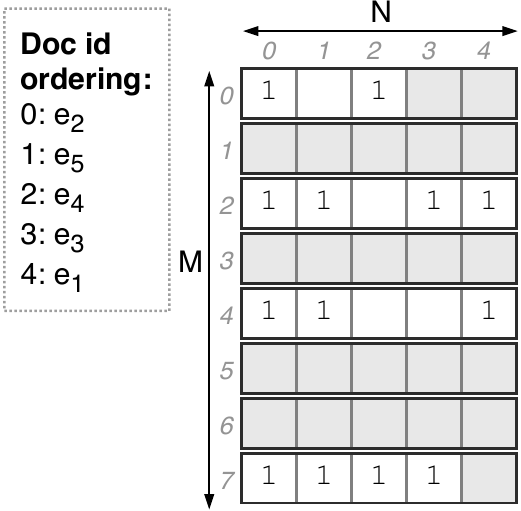}
    \caption{A Sparse Bloom Matrix constructed from Figure~\ref{fig:BM} by modifying the total ordering. The space cost reduces from 19 to 17 bits.}
    \label{fig:SBM}
\end{figure}

Bloom Matrix can be \textit{sparse}, where some or the bits are zero. This allows us to use a sparse storage method, to make its space cost less than $m\times N$. Furthermore, it is possible to further reduce the space cost by carefully selecting the total ordering so that sparse rows of the Bloom Matrix, have more zeros at the end, because trailing zeros can be spared by sparse vectors to save space. We name a Bloom Matrix with such ordering a \textit{Sparse Bloom Matrix}. We use the following example to illustrate the efficiency of Sparse Bloom Matrix over simple Bloom Matrix:

\begin{example}
    Take Example~\ref{BMexample} as an example. The Bloom Matrix in Figure~\ref{fig:BM} spares 19 bits (cells with grey background) according to the ordering in Example~\ref{BMexample}. In contrast, if the ordering is replaced by $\Pi'$ such that items are ordered by $\{e_2, e_5, e_4, e_3, e_1\}$, the new Sparse Bloom Matrix can be constructed as in Figure~\ref{fig:SBM}, which spares 2 more bits thanks to more zeros at the end of each vector.
\end{example}


The construct of a Sparse Bloom Matrix is straightforward: one can choose a total ordering $\Pi$ that sorts items in set $E$ in decreasing order of the number of assigned labels. This maximizes the probability of having more zeros at the end when using $\ENCODE$. More formally, let $C(e)$ be the number of labels assigned to an item $e$. We have to define the total ordering $\Pi$ so that $\Pi(e_i) > \Pi(e_j)$ iff $C(e_i) \leq C(e_j)$. Later in Section~\ref{experimental_results}, we will see that such ordering can archive in average 20\% reduction for space occupation.

\smallskip\noindent\textbf{SBM Complexity.}
In the case of a Sparse Bloom Matrix, the $\ADD$ and $\LOOKUP$ operations are the same. However, the initialization operation changes, because we need to compute the ordering $\Pi$. The speed of this operation depends on how the dataset is represented. If we represent it as an array of $|L|$ items, each item representing a $l \in L$ and containing a set of items of $E$, we need to scan the whole structure keeping a counter for each $e \in E$. The time would be therefore $\Theta(|L|N)$. If we represent it as an array of $N$ items, each item representing a $e \in E$ and containing a set of labels of $L$, and if each set has a precomputed size, the time would be $\Theta(N)$.

\section{Bloom Vector} \label{bloom_vector}



As we have seen in Section~\ref{preliminaries}, if we know a priori the number of items to be inserted in a Bloom Filter, we can choose its parameters so that the probability of \ac{FP} remains around a chosen value. Unlike Bloom Matrix, we aim for a new data structure with multiple variable length Bloom Filters, significantly reducing the memory overhead. We call this data structure  Bloom Vector. Therefore, we can expect Bloom Vector to acquire much less space by \ac{FP} rate than Bloom Matrix. We will see this in practice in Section~\ref{experimental_results}.


\subsection{Definition}

We define a Bloom Vector as a tuple $(G, \Pi)$, where $G$ is an \emph{associative array} of size $N = |E|$ in which each item corresponds to a Bloom Filter, and  $\Pi$ represents an ordering on the set $E$ as previously defined. As already stated, the Bloom Filters in $G$ can have different sizes and different hash functions from each other. 

\subsection{Operations}

Let us now define the operations on the Bloom Vector, which are based on the operations on Bloom Filter. 
\begin{algorithm}[h!]
    \caption{Bloom Vector: Add Operation}
    \begin{algorithmic}[1]
        \Procedure{Add}{$l$, $e$}
            \State $V \leftarrow \Call{Encode}{\Pi, e}$
            \State $I \leftarrow V.toList$
            \For {$i \leftarrow I$}
                \State $G[i].\Call{Add}{l}$
            \EndFor
        \EndProcedure
    \end{algorithmic}
    \label{algo:bvadd}
\end{algorithm}

\vspace{1mm}
\noindent\textbf{Add. } In order to add a label $l$ to the subset of $\mathcal{P}(E)$ given by $f(l)$, we compute $\ENCODE(\Pi, f(l))$. Let us call $I$ the indices of the bits set to 1 in $\ENCODE(\Pi, f(l))$. The add function on the Bloom vector $F$, is defined as:
\[ F.\ADD(l) := G[i].\ADD(l) \quad \forall i \in I \]

The add operation on Bloom Vector is illustrated in Alg.~\ref{algo:bvadd}. Suppose that we add the label $l$ to a  Bloom Vector. The $\ADD(l)$ operation executes only on the rows $V = \{v_1, ..., v_{|V|}\}$ corresponding to the bits set to 1 in the value given by $\ENCODE(\Pi, f(l))$. The bits to be set to 1 in each of those rows, however, are now determined by a new set of hash functions $H(l,m) = \{h_1(l,m), ..., h_k(l,m)\}$, where $m$ be the number of bits of the corresponding row. 

Noteworthy, Bloom Vector does not require having the same hash functions for each row (i.e., for each Bloom Filter), because either $\ADD$ or $\LOOKUP$ uses bits only a specified row decided by $\ENCODE$. However, considering the maintainability and the difficulty of implementation, one often like to use the same hash functions for all Bloom Filters. To achieve this goal, one needs to make an extension to the hash functions $H$ because it becomes it returns hash neighborhoods in a specific range, which becomes infeasible in a Bloom Vector where all Bloom Filters can have different lengths. As a solution, we can attach the maximal valid integer as a parameter of hash function, so that $H(l,m)=\{h_1(l,m),\cdots,h_k(l,m)\}$ returns $k$ the hash neighborhoods within the range $[0,m_k)$. Then, during $\ADD$ operation, we assign each $m$ as the number of bits in each row to ensure the output of $H(l,m)$ can always be mapped to a valid bit. In practice, this new function can be implemented easily by regulating the output of a hash algorithm $h(l)$ using its maximum possible value, e.g., $h(l,m)=\lfloor h(l)/2^{32}\times m \rfloor$ when $h(l)$ is an 32-bit MurmurHash.

\vspace{1mm}
\noindent\textbf{Lookup.} Alg.~\ref{algo:bvlookup} describes the lookup operation on Bloom Vectors. Similarly, in order to find out which subset of $\mathcal{P}(E)$ is labelled with $l$, let $V$ be a bitset defined, for $1 \leq i \leq N$ as:
\[
    V[i] :=
    \begin{cases}
        1 & \text{if } G[i].\LOOKUP(l) \\
        0 & \text{otherwise}
    \end{cases}
\]
The $\LOOKUP$ operation is defined as:
\[ F.\LOOKUP(l) := \DECODE(\Pi, V) \]

\begin{algorithm}[t]
    \caption{Bloom Vector: Lookup Operation}
    \begin{algorithmic}[1]
        \Procedure{Lookup}{$l$, $e$}
            \State Let $V$ be an empty bitset
            \For {$i \leftarrow [0,N)$}
                \If {$G[i].\Call{Lookup}{l}$}
                    \State $V[i] \leftarrow 1$
                \EndIf
            \EndFor
            \State \Return $\Call{Decode}{\Pi, V}$
        \EndProcedure
    \end{algorithmic}
    \label{algo:bvlookup}
\end{algorithm}

\begin{figure}[!h]
    \centering
    \includegraphics[width=0.8\linewidth]{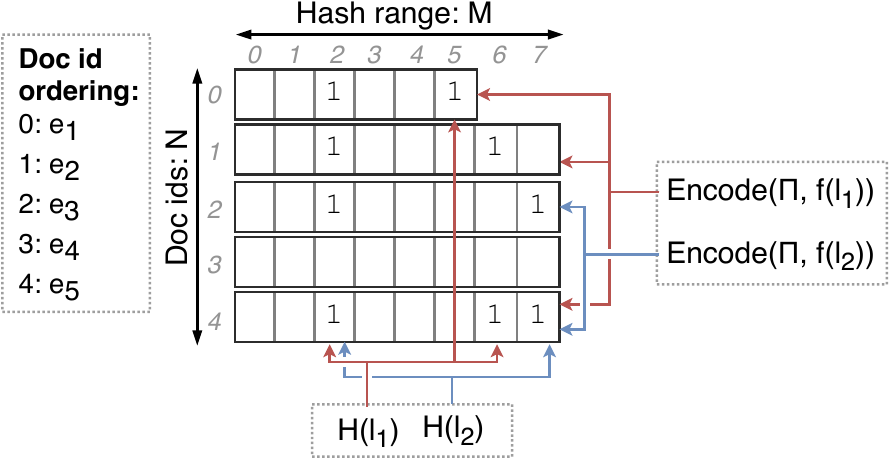}
    \caption{Example of a Bloom vector $\mathbf{G}[N,\_]$, obtained from Example \ref{BVexample}. The size of each Bloom Filter in the vector can be different. }
    \label{fig:BV}
\end{figure}

\begin{example} \label{BVexample}
    %

    Given 2 labels $L = \{l_1, l_2\}$, 5 items $E = \{e_1, \cdots, e_5\}$, an ordering $\Pi$ that sort items in input order, and two hash functions $H = \{h_1, h_2\}$ returning integers such that $H(l_1,6) = \{2,5\}$, $H(l_1,8) = \{2,6\}$, $H(l_2,6) = \{2,5\}$, and $H(l_2,8) = \{2,7\}$. The labels can be assumed as the words and items can be assumed as the documents.
    
    We now represent $f(l_1) = \{e_1, e_2, e_5\}$ and $f(l_2) = \{e_3, e_5\}$ with a Bloom Vector. As showing in Figure \ref{fig:BV}, 
    when adding $l_1$, $\ENCODE$ functions returns $\{1,1,0,0,1\}$ and hence we have three target rows $\mathbf{G}[1,\_]$, $\mathbf{G}[2,\_]$ and $\mathbf{G}[5,\_]$. Then, we perform a hash operation by having row length as input: $H(l_1,6) = \{2,5\}$ and $H(l_1,8) = \{2,6\}$ to obtain Columns 2 and 5 for Row 0, and Columns 2 and 6 for Rows 1 and 4. Finally, we need to set six bits: $\mathbf{G}[0,2]$, $\mathbf{G}[0,5]$, $\mathbf{G}[1,2]$, $\mathbf{G}[1,6]$, $\mathbf{G}[4,2]$, and $\mathbf{G}[4,6]$. $l_3$ can be added to the Bloom Vector by the same manner.
    
    When preforming $\LOOKUP(l_2)$, we first build a empty bitset $A$ with length $N$. We scan each row by checking the bit at position determined by the hash function. Specifically, for the first row, we get $H(l_2,6) = \{2,5\}$ to check the bits at Columns 2 and 5. Since both bits are $1$, we mark the \nth{1} bit of $A$ as $1$. For other rows, we get $H(l_2,8) = \{2,7\}$ to check Columns 2 and 7, and thereafter marks $A$'s \nth{3} and \nth{5} bits as $1$. Finally, we run $\DECODE$ on $A$ and get $\{e_1,e_3,e_5\}$, in which $e_1$ is a false positive.
\end{example}

\vspace{1mm}
\noindent\textbf{Multiple label lookup. }We can look up which items contain multiple labels by using the same algorithm for a single lookup, using in each Bloom Filter, the function for looking up multiple labels.



\subsection{False positive rate}

\begin{lemma}
    Given a set of labels $L=\{l_1,\cdots,l_{|L|}\}$ where each $l$ is associated with some items $e$. The average false positive rate of a Bloom Vector, when performing $\LOOKUP$ for all the labels in $L$, is 
    \begin{equation}\label{eq:BMVP_lemma}
        \overline{\text{FPR}} = \frac{\sum_{l \in L} \sum_{i \in I(l)} \left[1 - \left(1 - \frac{1}{m_i}\right)^{k_i n_i}\right]^{k_i}}{|L|}
    \end{equation}
    where $I(l)$ returns a set of row indices corresponding to positions of $1$'s in the output of $\ENCODE(l)$, $k_i$, $m_i$ and $n_i$ are the number of hash functions, the number of bits, and the number of labels added to $i$-th Bloom Filter, respectively.
\end{lemma}
\begin{proof}
    The $\LOOKUP$ operation on a Bloom Vector goes through each row to check whether the bits at columns given by $k$ hash functions are all $1$. Therefore, a false positive in this case is when all $k$ bits are being set to $1$ when adding other items into the structure.

    Recall Equation \ref{eq:bm_fp_2}. Given an arbitrary Bloom Filter with $m$ bits, $k$ hash functions, and $n$ added labels, its false positive rate is
    \begin{equation}
        \label{eq:bv_fp_1}
        (1-(1-\frac{1}{m})^{nk})^{k}
    \end{equation}

    Then, let $m_i$ and $k$ be the parameters of $i$-th Bloom Filter in a Bloom Vector, and let $n_i$ be the number of labels that the $i$-th Bloom Filter contains. We can use Equation \ref{eq:bv_fp_1} to derive the total expected \ac{FP} rate when looking up a label $l$:
    \begin{equation}\label{eq:BVFP2}
        \text{FPR} = \sum_{i \in I(l)} \left[1 - \left(1 - \frac{1}{m_i}\right)^{k_i n_i}\right]^{k_i}
    \end{equation}
    where $I(l)$ returns the index of Bloom Filters changed when adding $l$. Formally, it returns a set of row indices corresponding to positions of $1$'s in the output of $\ENCODE(l)$.

    Finally, given multiple labels $L$ for looking up, the average \ac{FP} rate is
    \begin{equation}\label{eq:BVFP}
        \overline{\text{FPR}} =\frac{\sum_{l \in L} \sum_{i \in I(l)} \left[1 - \left(1 - \frac{1}{m_i}\right)^{k_i n_i}\right]^{k_i}}{|L|}
    \end{equation}
\end{proof}


\subsection{Complexity}

Compared with Bloom Matrices, Bloom Vectors cost less space but perform slower lookups due to the traversal of all contained Bloom Vectors.

\begin{lemma}
    Bloom Vector has a space complexity $O(mN)$, $\ADD$ time complexity $O(|V|k)$ and $\LOOKUP$ complexity $\Theta(kN)$, where $m$ is the max size of hash neighborhood among all rows, $k$ is the max number of hash functions for each rows, $N$ is the \textit{total} number of items.
\end{lemma}

\begin{proof}
    Space: a Bloom Vector needs to store three components: (i) $m \times N$ in the worst case when all $N$ rows have an equal length $m$, (ii) The total ordering $\Pi$, which uses $2N$ space, and (iii) $kN$ hash functions. Therefore, the space cost is $O(mN)+2\Theta(N)+O(kN)=O(mN)$ when $k \ll m$. Note that this bound is not tight because different rows in the Bloom Vector can have different lengths, and hash functions can be reused for more than one rows if they have the same length.

    Time complexity: the $\ADD(l, e)$ operation needs to compute the ordering $V$ by executing $\ENCODE(\Pi, e)$, and then perform $\ADD$ operation to each row corresponding to $V$. Each $\ADD$ requires $O(k)$ time for hash functions. Therefore, the total time is $O(|V|) \cdot O(k) = O(|V|k)$ time, since time for updating one bit is negligible.

    A $\LOOKUP(l)$ operation needs to go through all rows. For each row, it needs to calculate $O(k)$ hash neighborhoods. Therefore, The total time is $N \cdot O(k) = O(kN)$.
\end{proof}

\vspace{1mm}
\noindent\textbf{Update.} Unlike Bloom Matrix, each Bloom Filter in a Bloom Vector has its own parameters. The $\ADD$ operation is performed on an individual filter. Therefore, it is possible to add new items incrementally to a vector and so the corresponding labels. Updating an already existing  Filter in a vector increases the FP rate. 







\section{Performance Evaluation} \label{experimental_results}

We chose Scala as the implementation language. We adapted the operations of both Bloom Multifilters to Scala, using auxiliary functions and taking advantage of the Map-Reduce paradigm. We evaluated the performance of our Bloom Multifilters with three different datasets. In this section, we first demonstrate the performance with synthetic datasets and then with a small real dataset used in various researches. The experiments were conducted on a machine with a quad-core processor at 2.3 GHz with eight logical processors, 16 GB of RAM at 1.6 GHz, and a 512 GB  SSD. 



\subsection{Dataset Generation}
In the first set of experiments, we used artificial datasets. This is useful to experiment with the behavior of Bloom Multifilters with different data distribution types. We implemented two functions. One function generates data having uniform distributions, and the other generates Zipf distributions. It is intuitive that Bloom Matrix is suitable for a dataset with Uniform distribution, as every Bloom Filter in the matrix are of equal size and so input sets. On the other hand, Bloom Vector is suitable for Zipf distribution, as the size of every Bloom Filter can be different depending on the size of the input sets. In Section~\ref{bloomtest}, we verify this with another set of experiments with a real dataset of unknown distribution.

\begin{table}[!h]
    \centering
    \begin{tabular}{| l | c | c |}
        \hline
        Name & Number of labels & File size \\
        \hline
        Uniform & $\sim 2500000$ & 12.2 MB \\
        Zipf & $\sim 30000$ & 171 kB \\
        \hline
    \end{tabular}
    \caption{Sizes of data sets used in the experiments.}
    \label{tab:datasets}
\end{table}

\vspace{1mm}
\noindent\textbf{Uniform.} The algorithm that generates uniform distributions, given $E$, $L$, and a probability $p$, for each $e \in E$, for each $l \in L$, decides with a probability of $p$ whether to assign such label $l$ to $e$ or not. We used  $p = 0.5$.  

\vspace{1mm}
\noindent\textbf{Zipf.} The algorithm that generates Zipf distributions, given $E$, $L$ and a real number $s$, generates the first $|E|$ Zipf rank numbers with exponent value $s$ and $N = |E|$, following the equation:
\begin{equation}
    f(k; s, n) = \frac{1 / k^s}{\sum_{i=1}^N (1/n^s)}
\end{equation}

\noindent  where $s = 0.8$. Then, for each $e_k \in E$, for each $l \in L$, the algorithm decides with a probability equal to the rank $k$ whether to assign such label $l$ to $e$ or not.

Table~\ref{tab:datasets} illustrates the properties of the synthetic datasets generated by the above methods, with $|E| = 500$ for both datasets. 
Both functions save the generated data into CSV files. Each row of the file contains the name of an element as the first value and the names of all the labels assigned to it following in the same row. This is similar to a row having a document id and the unique words in the document.




\subsection{Bloom Multifilters comparison}

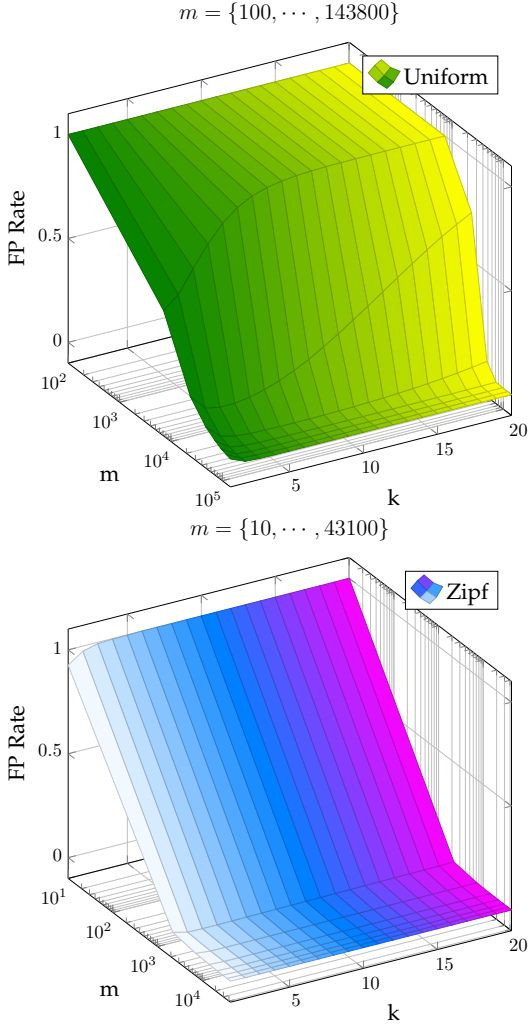
\begin{figure}[t]
\centering
\scalebox{0.7}{
\begin{tikzpicture}
\begin{axis}[view={60}{30},
		grid=both,
        legend pos=north east,
        legend cell align=left,
        width=10cm,height=10cm,
		xlabel=m,
		ylabel=k,
        title = {${m = \{100,\cdots,143800\}}$},
        zlabel=FP Rate, 
		label style={font=\large},
		title style={font=\large},
		legend style={font=\large},
        xmode = log,
        point meta={y*100},
		colormap/cool,
		no marks,
		]
\addplot3+[surf,colormap/greenyellow] file {tables/matrixuni.txt};\addlegendentry{Uniform}
\end{axis}
\end{tikzpicture}}

\scalebox{0.7}{
\begin{tikzpicture}
\begin{axis}[view={60}{30},
		grid=both,
        legend pos=north east,
        legend cell align=left,
        width=10cm,height=10cm,
		xlabel=m,
		ylabel=k,
        zlabel=FP Rate, 
		label style={font=\large},
		title style={font=\large},
		legend style={font=\large},
        xmode = log,
        title =  ${m = \{10,\cdots,43100\}}$,
        point meta={y*100},
		colormap/cool,
		no marks,
		]
\addplot3+[surf,colormap/cool] file {tables/matrixzipf.txt};\addlegendentry{Zipf}
\end{axis}
\end{tikzpicture}}
\caption{The performance of Bloom Multifilters with respect to the number of hash functions, the size of the Bloom Multifilters, and the distribution of labels. All the structures, BM, SBM, and BV, have similar characteristics.}
\label{fig:fpkmm}
\end{figure}

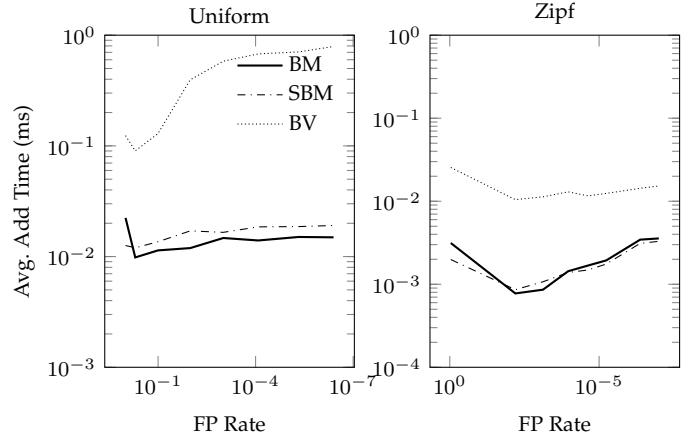
\begin{figure}[!h]
\centering
{\begin{tikzpicture}
\tikzstyle{every node}=[font=\footnotesize]
        \begin{groupplot}[
                width=4.9cm, 
        		height=6cm, 
        		title style={at={(0.5,0.96)},anchor=south,draw=none,fill=none},
        		group style={group size=2 by 1, horizontal sep=3em, vertical sep=3em}]
 
\nextgroupplot[
        legend cell align=left,
		title=Uniform,
        legend style={at={(0.03,1.0)},anchor=north west, draw=none, fill = none},
        legend pos = north east,
        xmode = log,
        ymode = log,
		ylabel= Avg. Add Time (ms),
        ylabel style = {yshift=-1ex},
    	ymin=0.001, ymax=1,
        x dir=reverse,
    	xlabel = FP Rate ,       
		]         
\addplot+[mark=none,draw=black, thick] table {

1	0.0223705600000000
0.503303710000000	0.00982617000000000
0.100612540000000	0.0113747500000000
0.0102491400000000	0.0119356600000000
0.000982180000000000	0.0147185600000000
8.27500000000000e-05	0.0139824200000000
4.80000000000000e-06	0.0150360400000000
4.00000000000000e-07	0.0149308900000000
};
\addplot+[mark=none,draw=black, dashdotted] table {
1	0.0125907100000000
0.503303710000000	0.0119985200000000
0.100612540000000	0.0136163800000000
0.0102491400000000	0.0170636200000000
0.000982180000000000	0.0165607400000000
8.27500000000000e-05	0.0185598000000000
4.80000000000000e-06	0.0186659200000000
4.00000000000000e-07	0.0190895100000000
};
\addplot+[mark=none,draw=black,  densely dotted] table {
1	0.123797050000000
0.503303710000000	0.0901372900000000
0.100612540000000	0.129033180000000
0.0102491400000000	0.395462010000000
0.000982180000000000	0.582535990000000
8.27500000000000e-05	0.677675860000000
4.80000000000000e-06	0.705585810000000
4.00000000000000e-07	0.790534740000000
};
\legend{BM, SBM, BV};
\nextgroupplot[
        legend cell align=left,
		title=Zipf,
        legend style={at={(0.2,1.0)},anchor=north east, draw=none, fill = none},
        legend pos = north east,
        xmode = log,
        ymode = log,
        ylabel style = {yshift=-1ex},
    	ymin=0.0001, ymax=1,
        x dir=reverse,
    	xlabel = FP Rate ,       
		]         
\addplot+[mark=none,draw=black, thick] table {

0.926336710000000	0.00313870000000000
0.00636882000000000	0.000775260000000000
0.000755510000000000	0.000860650000000000
0.000108660000000000	0.00143578000000000
2.33900000000000e-05	0.00168192000000000
5.77000000000000e-06	0.00193725000000000
4.20000000000000e-07	0.00344606000000000
1.00000000000000e-07	0.00356481000000000
};

\addplot+[mark=none,draw=black, dashdotted] table {
0.926336710000000	0.00198654000000000
0.00636882000000000	0.000860680000000000
0.000755510000000000	0.00107178000000000
0.000108660000000000	0.00138633000000000
2.33900000000000e-05	0.00148602000000000
5.77000000000000e-06	0.00175174000000000
4.20000000000000e-07	0.00310836000000000
1.00000000000000e-07	0.00329043000000000
};
\addplot+[mark=none,draw=black, densely dotted] table {
0.926336710000000	0.0255093100000000
0.00636882000000000	0.0104521200000000
0.000755510000000000	0.0113020100000000
0.000108660000000000	0.0129749400000000
2.33900000000000e-05	0.0116077800000000
5.77000000000000e-06	0.0124317600000000
4.20000000000000e-07	0.0143169200000000
1.00000000000000e-07	0.0152793600000000
};
\end{groupplot}
\end{tikzpicture}}
\vspace{-3mm}
\caption{Time required to execute an $\ADD$ operation or to add a label to Bloom Multifilters by FP rate, using optimal $m$ and $k$ for each point.}
\label{fig:addtime}
\end{figure}



\begin{figure}[t]
\centering
{\begin{tikzpicture}
\tikzstyle{every node}=[font=\scriptsize]
        \begin{groupplot}[
                width=4.8cm, 
        		height=6cm, 
        		title style={at={(0.5,0.96)},anchor=south,draw=none,fill=none},
        		group style={group size=2 by 1, horizontal sep=3em, vertical sep=3em}]
 
\nextgroupplot[
        legend cell align=left,
		title={Uniform},
        legend style={at={(0.03,1.0)},anchor=north west, draw=none, fill = none},
        legend pos = north east,
        xmode = log,
		ylabel= Memory Overhead (MB),
        ylabel style = {yshift=-3ex},
    	ymin=0.01, ymax=16,
        x dir=reverse,
    	xlabel = FP Rate , 
		]        
\addplot+[mark=none,draw=black, thick] table {
0.90000000 0.07608
0.50000000 0.45808
0.10000000  1.50
0.01000000  3.00
0.00100000  4.50
0.00010000  5.99
0.00001000  7.49
0.00000100  8.99
};
\addplot+[mark=none,draw=black, dashdotted] table {
0.90000000  0.07594
0.50000000 0.34438
0.10000000  1.07
0.01000000  2.31
0.00100000  3.38
0.00010000  4.44
0.00001000  5.67
0.00000100  6.73
};
\addplot+[mark=none,draw=black, densely dotted] table {
0.90000000  0.07806
0.50000000 0.46013
0.10000000  1.51
0.01000000  3.00
0.00100000  4.50
0.00010000  6.00
0.00001000  7.49
0.00000100  8.99
};

\addplot+[mark=none,draw=black, dashed] table {
0.90000000  13.99
0.50000000  13.99
0.10000000  13.99
0.01000000  13.99
0.00100000  13.99
0.00010000  13.99
0.00001000  13.99
0.00000100  13.99
};
\legend{BM, SBM, BV}
\nextgroupplot[
        legend cell align=left,
		title={Zipf},
        legend style={at={(0.03,1.0)},anchor=north west, draw=none, fill = none},
        legend pos = south east,
        xmode = log,
        ymode = log,
        ylabel style = {yshift=-1ex},
    	ymin=0.001, ymax=10,
        x dir=reverse,
    	xlabel = FP Rate ,       
		]         

\addplot+[mark=none,draw=black, thick] table {
0.926336710000000	0.00820112500000000
0.00636882000000000	0.145076125000000
0.000755510000000000	0.457576125000000
0.000108660000000000	0.907576125000000
2.33900000000000e-05	1.35757612500000
5.77000000000000e-06	1.80757612500000
4.20000000000000e-07	2.25132612500000
1.00000000000000e-07	2.70132612500000
};
\addplot+[mark=none,draw=black, dashdotted] table {
0.926336710000000	0.00819975000000000
0.00636882000000000	0.133163625000000
0.000755510000000000	0.368822125000000
0.000108660000000000	0.646884125000000
2.33900000000000e-05	0.893867125000000
5.77000000000000e-06	1.12966337500000
4.20000000000000e-07	1.61816087500000
1.00000000000000e-07	1.87803725000000
};
\addplot+[mark=none,draw=black, densely dotted] table {
1	0.0103948750000000
0.505681690000000	0.0149928750000000
0.103238140000000	0.0275748750000000
0.0103415700000000	0.0455781250000000
0.00104888000000000	0.0635826250000000
0.000100380000000000	0.0815842500000000
1.14300000000000e-05	0.0995846250000000
1.47000000000000e-06	0.117590000000000
4.20000000000000e-07	0.13
1.00000000000000e-07	0.15
};

\addplot+[mark=none,draw=black, dashed] table {
1	0.30376
0.505681690000000	0.30376
0.103238140000000	0.30376
0.0103415700000000	0.30376
0.00104888000000000	0.30376
0.000100380000000000	0.30376
1.14300000000000e-05	0.30376
1.47000000000000e-06	0.30376
4.20000000000000e-07	0.30376
1.00000000000000e-07	0.30376
};

\end{groupplot}
\end{tikzpicture}}
\caption{Memory usage by FP rate, using optimal $m$ and $k$ for each point. The horizontal dashed line represents memory usage by conventional data structures representing the data.}
\label{fig:memovh}
\end{figure}
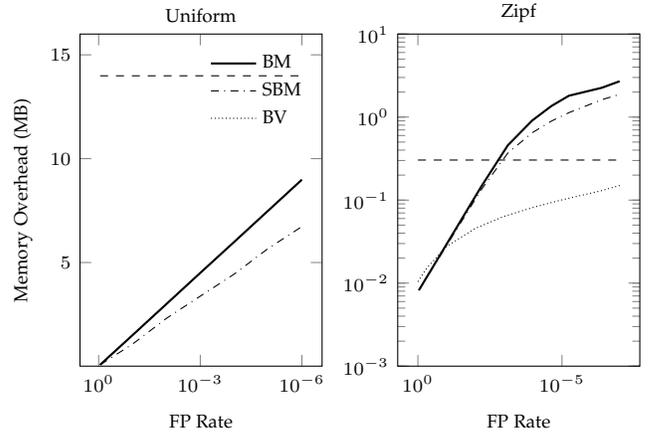

We evaluate and compare the performance of our Bloom Multifilters, Bloom Matrix (BM), Spare Bloom Matrix (SBM), and Bloom Vector (BV), with above two datasets. We first investigate their FP rates with respect to the number of hash functions, $k$, and their sizes, $m$. Therefore, we construct the Bloom Multifilters with different combinations of $m$ and $k$. Next, we compare their performance for memory overhead, $\ADD$, and $\LOOKUP$ operations with respect to various FP rates. 

\vspace{1mm}
\noindent\textbf{False Positive Rates.} 
As we can see in Figure~\ref{fig:fpkmm}, the \ac{FP} rate decreases as $k$ increases, however, increases again after a certain number of hash functions, as expected from our theoretical analysis in the earlier sections. All the Bloom Multifilters perform the same.  

The FP rates of BM and SBM also decrease as $m$ increases for both Uniform and Zipf distributions as shown in   Figure~\ref{fig:fpkmm}. With Zipf distribution, they can achieve lower FP rates with smaller sizes. It is difficult to demonstrate such relations for a Bloom Vector when the size of each Bloom Filter is different in the vector. Nevertheless, if all the vectors are of equal size and have the same number of hash functions, then  Bloom Vector also depicts exactly the same performance. In this case, for every combination of $m$ and $k$, all the Bloom Filters in the Vectors are of equal size.


\vspace{1mm}
\noindent\textbf{$\ADD$ Operation Times Vs FP Rates.} In Figure~\ref{fig:addtime}, we notice that the $\ADD$ operation takes near a linear time on Bloom Matrix with Uniform distribution. This is expected as an $\ADD$ operation costs $O(N \log |e|)$. The operation time on Bloom Vector is linear too with Zipf distribution. However, the $\ADD$  operation takes more time on Bloom Vector than Bloom Matrix with both distributions. This is because the time depends on the size of the vector that it uses to store the result of $\ENCODE$ operation, i.e., $V \leftarrow \Call{Encode}{\Pi, e}$,  multiplied by $k$ ($O(|e| + |V|k)$).

\vspace{1mm}
\noindent\textbf{Memory Overhead Vs FP Rate.} Figure~\ref{fig:memovh} demonstrates the memory overhead of the structures as the FP rate decreases. 
We notice that all the Bloom Multifilters are well below the horizontal line when representing the dataset of Uniform distribution. Bloom Vector occupies a similar space to its basic counterpart, i.e., the Bloom Matrix. This is because all the Bloom Filters of Bloom Vectors are of the same size; therefore, both structures have similar parameters. On the other hand, the Sparse Bloom Matrix occupies the least space; the Sparse Bloom Matrix sets the ordering on the set $E$ to maximize the number of zeros at the end of each row. This makes it the most suitable Bloom Multifilter to represent uniformly distributed data if our goal is to spare as much space as possible.

On the other hand, with Zipf distributed dataset, both Bloom Matrix, and Sparse Bloom Matrix perform poorly concerning space. Although the Sparse Bloom Matrix performs better than Bloom Matrix, they both are above the dashed line for a specific FP rate. Bloom Vector is the most space efficient with a Zipf distribution and remains below the dashed line even with an FP rate of $10^{-6}$. The explanation is simple: Bloom Vector uses different optimized sizes for each row, i.e., $m$ is different for different Bloom Filters, where each row represents the labels associated with a particular element $e\in E$. Therefore, with sparse rows of the distribution, it does not waste space as the other Bloom Multifilters do.


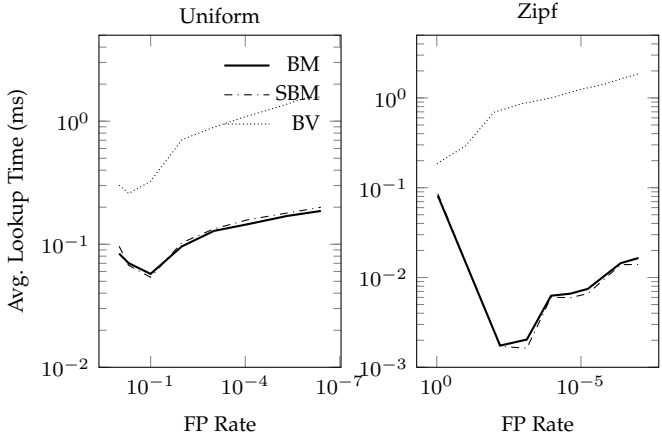
\begin{figure}[t]
\centering
{\begin{tikzpicture}
\tikzstyle{every node}=[font=\footnotesize]
        \begin{groupplot}[
                width=4.8cm, 
        		height=6cm, 
        		title style={at={(0.5,0.96)},anchor=south,draw=none,fill=none},
        		group style={group size=2 by 1, horizontal sep=3em, vertical sep=3em}]
 
\nextgroupplot[
        legend cell align=right,
		title={Uniform},
        legend style={at={(0.06,1.0)},anchor=north west, draw=none, fill = none},
        legend pos = north east,
        xmode = log,
        ymode = log,
		ylabel= Avg. Lookup Time (ms),
        ylabel style = {yshift=-1ex},
    	ymin=0.01, ymax=5,
        x dir=reverse,
    	xlabel = FP Rate ,       
		]         
\addplot+[mark=none,draw=black, thick] table {
1	0.0839117500000000
0.503303710000000	0.0703212200000000
0.100612540000000	0.0573873300000000
0.0102491400000000	0.0960679300000000
0.000982180000000000	0.128317580000000
8.27500000000000e-05	0.145502110000000
4.80000000000000e-06	0.169595180000000
4.00000000000000e-07	0.186417970000000
};
\addplot+[mark=none,draw=black,  dashdotted] table {
1	0.0963992800000000
0.503303710000000	0.0674818900000000
0.100612540000000	0.0538890500000000
0.0102491400000000	0.102901500000000
0.000982180000000000	0.133394180000000
8.27500000000000e-05	0.158114410000000
4.80000000000000e-06	0.179072690000000
4.00000000000000e-07	0.199821420000000
};
\addplot+[mark=none,draw=black, densely dotted] table {
1	0.302040390000000
0.500513850000000	0.258727540000000
0.101278510000000	0.324205940000000
0.0100168800000000	0.708631950000000
0.00101016000000000	0.890762650000000
0.000101140000000000	1.08629352000000
7.60000000000000e-06	1.32358513000000
1.20000000000000e-06	1.54748751000000
4.20000000000000e-07	1.57
};
\legend{BM, SBM, BV}
\nextgroupplot[
        legend cell align=left,
		title={Zipf},
        legend style={at={(0.03,1.0)},anchor=north west, draw=none, fill = none},
        legend pos = north east,
        xmode = log,
        ymode = log,
        ylabel style = {yshift=-1ex},
    	ymin=0.001, ymax=5,
        x dir=reverse,
    	xlabel = FP Rate ,       
		]         
\addplot+[mark=none,draw=black, thick] table {
0.926336710000000	0.0817162600000000
0.00636882000000000	0.00174354000000000
0.000755510000000000	0.00203724000000000
0.000108660000000000	0.00626900000000000
2.33900000000000e-05	0.00660602000000000
5.77000000000000e-06	0.00746804000000000
4.20000000000000e-07	0.0144386900000000
1.00000000000000e-07	0.0164880600000000
};

\addplot+[mark=none,draw=black, dashdotted] table {
0.926336710000000	0.0856207700000000
0.00636882000000000	0.00170871000000000
0.000755510000000000	0.00162842000000000
0.000108660000000000	0.00602883000000000
2.33900000000000e-05	0.00595380000000000
5.77000000000000e-06	0.00662075000000000
4.20000000000000e-07	0.0139113400000000
1.00000000000000e-07	0.0139638200000000
};

\addplot+[mark=none,draw=black, densely dotted] table {
1	0.185270100000000
0.505681690000000	0.211503450000000
0.103238140000000	0.290816360000000
0.0103415700000000	0.693747090000000
0.00104888000000000	0.865743760000000
0.000100380000000000	1.00537148000000
1.14300000000000e-05	1.23251487000000
1.47000000000000e-06	1.42848582000000
4.20000000000000e-07	1.63
1.00000000000000e-07	1.86
};
\end{groupplot}
\end{tikzpicture}}
\caption{Lookup time of a single label on Bloom Multifilters with different FP rates. We used optimal $m$ and $k$ for each point.}
\label{fig:lookuptime}
\end{figure}


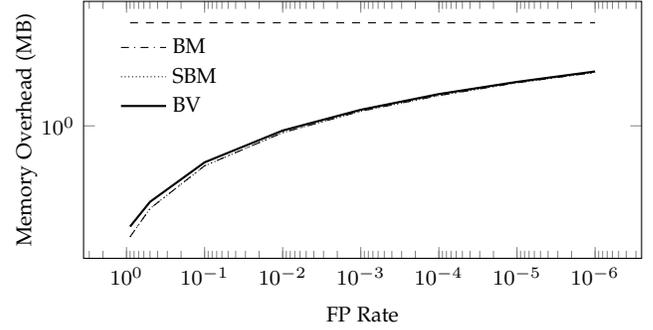
\begin{figure}[!h]
\centering
{\begin{tikzpicture}
\tikzstyle{every node}=[font=\footnotesize]
        \begin{groupplot}[
                width=9cm, 
        		height=5cm, 
        		title style={at={(0.5,0.96)},anchor=south,draw=none,fill=none},
        		group style={group size=2 by 1, horizontal sep=3em, vertical sep=3em}]
 
\nextgroupplot[
        legend cell align=left,
        legend style={at={(0.05,0.9)},anchor=north west, draw=none, fill = none},
        xmode = log,
        ymode = log,
		ylabel= Memory Overhead (MB),
        ylabel style = {yshift=-3ex},
        x dir=reverse,
    	xlabel = FP Rate , 
		]        
\addplot+[mark=none,draw=black, dashdotted] table {
0.90000000	0.163
0.50000000	0.259
0.10000000	0.521
0.01000000	0.895
0.00100000	 1.270
0.00010000	 1.640
0.00001000	 2.020
0.00000100	 2.390
};
\addplot+[mark=none,draw=black, densely dotted] table {
0.90000000	0.163
0.50000000	0.259
0.10000000	0.520
0.01000000	0.895
0.00100000	 1.270
0.00010000	 1.640
0.00001000	 2.020
0.00000100	 2.390
};
\addplot+[mark=none,draw=black, thick] table {
0.90000000	0.193
0.50000000	0.289
0.10000000	0.551
0.01000000	0.926
0.00100000	 1.300
0.00010000	 1.680
0.00001000	 2.050
0.00000100	 2.430
};

\addplot+[mark=none,draw=black, dashed] table {
0.90000000  5.380
0.50000000  5.380
0.10000000  5.380
0.01000000  5.380
0.00100000  5.380
0.00010000  5.380
0.00001000  5.380
0.00000100  5.380
};
\legend{BM, SBM, BV}

\end{groupplot}
\end{tikzpicture}}
\caption{Memory usage by FP rate. The horizontal dashed line represents memory usage by conventional data structures representing 20ng-test-stemmed data.}
\label{fig:rmemovh}
\end{figure}

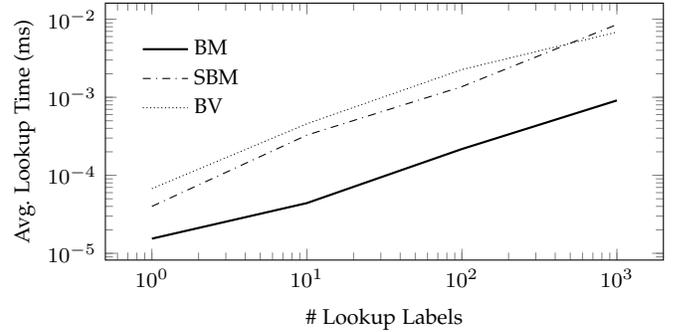
\begin{figure}[t]
\centering
{\begin{tikzpicture}
\tikzstyle{every node}=[font=\footnotesize]
        \begin{groupplot}[
                width=9cm, 
        		height=5cm, 
        		title style={at={(0.5,.96)},anchor=south,draw=none,fill=none},
        		group style={group size=2 by 1, horizontal sep=3em, vertical sep=4em}]

 \nextgroupplot[
        legend cell align=left,
        legend style={at={(0.05,0.9)},anchor=north west,draw=none,fill=none},
        legend columns=1,
        xmode = log,
        ymode = log,
		ylabel= Avg. Lookup Time (ms),
        ylabel style = {yshift=-1ex},
    	xlabel = \# Lookup Labels ,       
		]  
 \addplot+[mark=none,draw=black, thick] table {
1	0.00001542
10	0.00004405
100	0.00021715
1000 0.00091132	
};
\addplot+[mark=none,draw=black, dashdotted] table {
1	0.00003994
10	0.00032733
100	0.00136881
1000 0.00855334
};

\addplot+[mark=none,draw=black, densely dotted] table {
1 0.00006752
10 0.00045751
100 0.00226477
1000 0.00683597
};
\legend{BM, SBM, BV};
\end{groupplot}

\end{tikzpicture}}
\caption{Lookup time for one and multiple labels from the Bloom Multifilters at an expected FP rate of 0.1.}
\label{fig:multilookup}
\end{figure}

\vspace{1mm}
\noindent\textbf{$\LOOKUP$ Operation Times Vs FP Rates.} 
Once the Bloom Multifilters are constructed, we perform $\LOOKUP$ operation on the structures.  In Figure~\ref{fig:lookuptime}, we can see that Both Bloom Matrix and Sparse Bloom Matrix have a similar lookup time, and they are faster than Bloom Vector in all situations. This is because the lookup time of the Bloom Matrix depends only on $k$, ($\Theta(k)$). The Bloom Vector takes much more time and has a linear time too ($O(Nk + |V|)$). This might be because the $\LOOKUP$ operation requires to iterate of a $E$ number of Bloom Filters in the Vector, and ends up having an increasing workload. 




Sometimes it is crucial to find labels sharing multiple items. Instead of iterating over a list of labels, search them together. 
Multiple label lookup with our Bloom Multifilters allows us to find out the sets to which multiple labels belong in common without doing a lookup of each label. We present the multiple label lookup with real datasets in the next section.


\subsection{Discussion}




Our Bloom Multifilters obey the principle of standard Bloom Filter. Their space efficiency depends on the distribution of the labels in the input dataset and Bloom Vector is the most space efficient with Zipf distributed data. The performance of basic $\ADD$ and $\LOOKUP$ depends on their construction, and they are faster on Bloom Matrix,  similar to the theoretical analyses  presented in Section 4. 

\section{Bloom Test}
\label{bloomtest}

In this section, we test our Bloom Multifilters with a real dataset with unknown distribution, and illustrate the procedure to find the best Bloom Multifilter through small and fast empirical experiments. Specifically, we used the \textsf{20ng-test-stemmed} corpus, which is the 20 Newsgroups corpus test dataset with stemmed words, from the datasets for single-label text categorization \cite{Car07}. The dataset has $|E|$ = 7527 documents and $|L|$ = 625635 words. In this case, we do not have prior knowledge about the distribution of the dataset; i.e., Uniform or Zipf. We construct the Bloom Multifilters around certain Expected False Positive rates, i.e., $p\in \{0.9, 0.5, 10^{-1}, 10^{-2},\cdots, 10^{-6}\}$. The value of $m$ and $k$ were computed according to equations \ref{mbf} and \ref{kbf}. In the case of Bloom Matrix and Sparse Bloom Matrix, we compute $m$ based on the average amount of words in each document, i.e., $m =-\frac{|L|}{|N|}\cdot \frac{\ln p}{\ln^2{2}}$. Next, we compute $k$ according to equation 2b. On the other hand, for Bloom Vector, we compute $m$ and $k$ for every document in  $E$ using equation \ref{mbf} and \ref{kbf}.


\vspace{1mm}
\noindent \textbf{Performance.} 
Figure~\ref{fig:rmemovh} compares the memory overhead of the structures with the real dataset. We notice that all the structures perform similarly and even with $10^{-6}$ FP rate the occupy 45\% of the actual size. 
Figure~\ref{fig:multilookup} shows the performance of $\LOOKUP$ operation with multiple labels for given FP rates. We notice that the number of labels has a negligible effect on lookup time. As we discussed before, this is because the only additional workload for multiple labels is that the lookup function needs to compute the hash neighborhood of all the labels. The lookup time increases as the FP rate decreases, however, Bloom Matrix seems to be both space efficient and the lookup operation is also the time efficient.  

Nevertheless, the $\LOOKUP$ operation works with a complete set of queries. In other words, all the labels are searched together by $\land$ operation.  Performing an $\lor$ operation, i.e., $V \leftarrow \lor \wedge \mathbf{G}[i,\_]$, on Bloom Matrix for subset queries will provide invalid indexes of the items while performing the $\Call{Decode}{\Pi, V}$ operation, for example. In order to find the match for such subset of the queries, it requires to look for the items of each label and then perform OR operation, i.e., after the $\Call{Decode}{\Pi, V}$ operation.

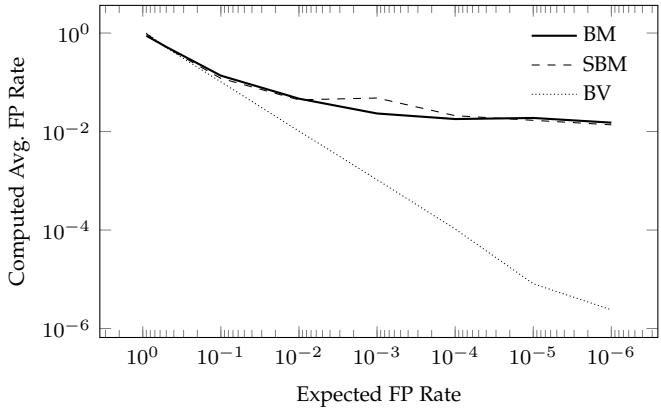
\begin{figure}[t]
\centering
{\begin{tikzpicture}
\tikzstyle{every node}=[font=\footnotesize]
        \begin{groupplot}[
                width=9cm, 
        		height=6cm, 
        		title style={at={(0.5,0.96)},anchor=south,draw=none,fill=none},
        		group style={group size=2 by 1, horizontal sep=3em, vertical sep=3em}]
 
\nextgroupplot[
        legend cell align=left,
        legend style={at={(0.03,1.0)},anchor=north west, draw=none, fill = none},
        legend pos = north east,
        xmode = log,
        ymode = log,
		ylabel= Computed Avg. FP Rate,
        ylabel style = {yshift=-1ex},
        x dir=reverse,
    	xlabel = Expected FP Rate , 
		]        
\addplot+[mark=none,draw=black, thick] table {
0.90000000	0.88304096
0.50000000	0.52311987
0.10000000	0.13646027
0.01000000	0.04650545
0.00100000	0.02325272
0.00010000	0.0179402
0.00001000	0.01886792
0.00000100	0.01514749
};
\addplot+[mark=none,draw=black, dashed] table {
0.90000000	0.96664894
0.50000000	0.51102841
0.10000000	0.12014429
0.01000000	0.04399841
0.00100000	0.04796705
0.00010000	0.02099389
0.00001000	0.01688381
0.00000100	0.01369135
};

\addplot+[mark=none,draw=black, densely dotted] table {
0.90000000	0.99866825
0.50000000	0.50327551
0.10000000	0.10192360
0.01000000	0.01021684
0.00100000	0.00105126
0.00010000	0.00010488
0.00001000	0.00000812
0.00000100	0.00000240
};
\legend{BM, SBM, BV}

\end{groupplot}
\end{tikzpicture}}
\caption{Computed avg. FP rate vs Expected FP rates for \textsf{20ng-test-stemmed} dataset (unknown distribution). The avg. FP rates were computed for a $\LOOKUP$ operation with 1000 labels.}
\label{fig:fpratio}
\end{figure}

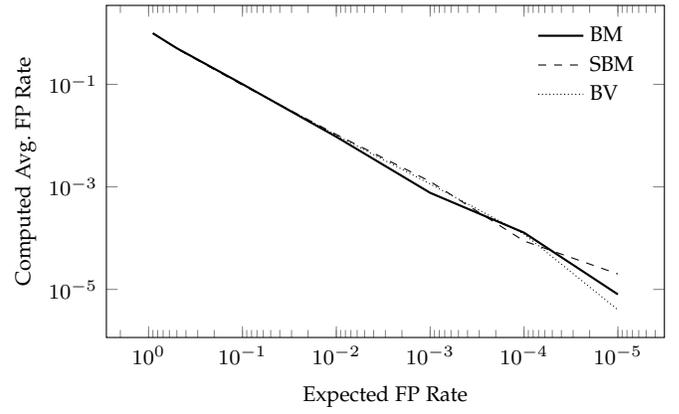
\begin{figure}[!h]
\centering
{\begin{tikzpicture}
\tikzstyle{every node}=[font=\footnotesize]
        \begin{groupplot}[
                width=9cm, 
        		height=6cm, 
        		title style={at={(0.5,0.96)},anchor=south,draw=none,fill=none},
        		group style={group size=2 by 1, horizontal sep=3em, vertical sep=3em}]
 
\nextgroupplot[
        legend cell align=left,
        legend style={at={(0.03,1.0)},anchor=north west, draw=none, fill = none},
        legend pos = north east,
        xmode = log,
        ymode = log,
		ylabel= Computed Avg. FP Rate,
        ylabel style = {yshift=-1ex},
        x dir=reverse,
    	xlabel = Expected FP Rate , 
		]        
\addplot+[mark=none,draw=black, thick] table {
0.90000000	0.98971426
0.50000000	0.50583744
0.10000000	0.10092603
0.01000000	0.00947977
0.00100000	0.00076518
0.00010000	0.0001279
0.00001000	0.00000799
0.00000100	0
};
\addplot+[mark=none,draw=black, dashed] table {
0.90000000 	0.99134022
0.50000000	0.49471614
0.10000000	0.0965404
0.01000000	0.010501
0.00100000	0.00127425
0.00010000	0.00008798
0.00001000	0.00001999
0.00000100	0
};

\addplot+[mark=none,draw=black, densely dotted] table {
0.90000000	1
0.50000000	0.50050926
0.10000000	0.10037906
0.01000000	0.01015928
0.00100000	0.00114038
0.00010000	0.00012013
0.00001000	0.000004
0.00000100	0
};
\legend{BM, SBM, BV}

\end{groupplot}
\end{tikzpicture}}
\caption{Computed avg. FP rate vs Expected FP rates for the synthetic data (Uniform distribution). The avg. FP rates were computed for a $\LOOKUP$ operation with 1000 labels.}
\label{fig:unifpratio}
\end{figure}


\noindent\textbf{Bloom Test.} We next evaluate the performance of the structures in terms of FP rate. While we construct them with expected FP rates, we also compute the FP rates for the lookup operations performed in this section. Figure~\ref{fig:fpratio} compares the expected FP rate with the computed FP rate. We notice that only the computed FP rate of Bloom vector follow the expected FP rate, whereas the FP rates of Bloom matrix and Sparse Bloom Matrix always higher than $10^{-2}$. Although such FP rates should be sufficient for many applications, this behavior of Bloom Matrix also exposes some characteristic about the input data. In other words, the labels in \textsf{20ng-test-stemmed} are not uniformly distributed among the items. In order to verify this, we further performed similar experiments with the uniformly distributed synthetic data. We constructed all three Bloom Multifilters with the mentioned expected FP rates. Figure~\ref{fig:unifpratio} shows that the FP rates for the $\LOOKUP$ operations on them follow the expected FP rates of the structures. The FP rates were computed according to the following:

\begin{equation}
 \text{FPR} = \frac{(|f(l)|-|e|)} {(|E|-|f(l)|)+(|f(l)|-|e|)}
\end{equation} 
where $(|E|-|f(l)|)$ is the true negative and  $(|f(l)|-|e|)$ is the false positive. $f(l)$ is the number of items returned by the $\LOOKUP$ function and $|e|$ is the actual number of items containing label $l$ in the input sets.

The above finding suggests that Bloom Matrix can be used to indicate whether an input dataset, with multiple sets, follows Uniform distribution or not. Since the FP rates of all the structures follow the expected rates until $10^{-2}$,  a simple test can first construct a Bloom Matrix with an expected  FP rate $10^{-3}$ and then compare with the FP rate of a lookup operation. If the difference is significant, the test can conclude that the input data distribution is not Uniform. Therefore, the test can be used to decide, based on desired and real FP rates, which structure we should choose for having an FP rate close to what we want. With all the required computation for $\ADD$ operation and $\LOOKUP$ operation for 1000 labels, the test requires only 1.3 MB space and 12 seconds for the \textsf{20ng-test-stemmed} dataset with the mentioned hardware.

Nevertheless, the unexpected performance of Bloom Matrix stems from the fact that it has to rely on average size of the input sets for unknown distribution, i.e., $\frac{|L|}{|E|}$. In other words, the sizes of the corresponding Bloom Filters estimated as $m =-\frac{|L|}{|N|}\cdot \frac{\ln p}{\ln^2{2}}$ are too small compared to the actual size of the input sets.   

\section{Related work}
\label{relatedwork}

There are many extensions of the Bloom Filter. In this section, we discuss some of them.

\vspace{1mm}
\noindent\textbf{Counting Bloom Filter.} The Counting Bloom Filter \cite{FCA00} is a variant of standard Bloom Filter that allows a delete operation without creating the chance of \acp{FN}. However, instead of using a bitset, it uses an array of integers. The $\ADD$ operation increments the integers to which a label is mapped to with the hash functions, while the delete operation decrements them. The lookup function simply checks whether all of the integers to which a label is mapped to are higher than 0. Stateful  Bloom Filter~\cite{BMP06}  also extends standard Bloom filter. The stored elements are neither bits nor counters, rather each element of $m$ represents a value corresponding to a state, i.e., state, and state counter, for identifying P2P traffic and congestion control for video traffic.

\vspace{1mm}
\noindent\textbf{Compressed Bloom Filter. }The Compressed Bloom Filter \cite{Mit02} was proposed to reduce the number of bits broadcast in network applications, \ac{FP} rate, and lookup time. This advantage comes at the cost of introducing a processing time for compression and decompression. The compression algorithm proposed in the original work \cite{Mit02} is Arithmetic Coding \cite{MNW98}, which is a lossless data compression technique. Unlike standard Bloom Filter,  the optimal $k$ is chosen to optimize the result of the compression algorithm in a Compressed Bloom Filter. This results in a choice of $k$ lower than in a standard Bloom Filter.


\vspace{1mm}
\noindent\textbf{Split Bloom Filter.} The Split Bloom Filter \cite{CFL04} uses a bitset split in multiple bins. Each bin has an associated hash function, and the hash functions are all different from each other. Whenever an element is added, it is added to all bins.

More formally, a Split Bloom Filter is composed by $k$ bins $G = \{B_1, ..., B_k\}$ each having size $m$, where $k$ is also the number of hash functions. Each hash function $h_i(x)$ is associated to the bitset $B_i$ having the same index. Whenever an element $x$ is added, we set to 1 the bits $B_i[h_i(x)]$, for $1 \leq i \leq k$.

The lookup operation works similarly, but it checks whether all bits $B_i[h_i(x)]$, for $1 \leq i \leq k$, are set to 1.


\vspace{1mm}
\noindent\textbf{Scalable/Dynamic Bloom Filter}
A Scalable Bloom Filter~\cite{ABP07} starts with a Split Bloom Filter \cite{CFL04} with $k_0$ bins and $P_0$ expected \ac{FP} rate, which can support at most a number of elements that keep the \ac{FP} rate below $P_0$. When the filter gets full, another one is added with $k_1$ bins and $P_1 = P_0 r$ expected \ac{FP} rate, where $r$ is a tightening ratio decided during the implementation. This is useful when the number of labels in a set is unknown. Alternatively, dynamic Bloom Filters~\cite{GWC10} can be used as the size of the data grows with time.









\vspace{1mm}
\noindent\textbf{Bloomier Filter.} The Bloom Filter can encode only Boolean functions. The Bloomier Filter \cite{CKR04} was proposed to represent arbitrary functions on finite sets.

Let $E = \{e_1, ..., e_N\}$ and $R = \{1, ..., |R| - 1\}$. Let $A = \{(e_1, v_1), ..., (e_N, v_N)\}$ be an assignment, where $v_i \in R$ for $1 \leq i \leq N$. The encoding of such assignment also be seen as a function $f : E \rightarrow R$ defined as:
\[
    f(x) =
    \begin{cases}
        v_i & \text{if } x \in E \\
        \varnothing & \text{otherwise}
    \end{cases}
\]

The Bloomier filter uses a bit matrix to encode the function previously defined. In order to build such a matrix, it uses a non-trivial algorithm, which can be found in the original work \cite{CKR04}. In this algorithm, the Bloomier Filter uses two functions called $\ENCODE$ and $\DECODE$. In Section~\ref{problem} we define two similar functions that we call with the same names, which our Bloom Multifilters use.

\vspace{1mm}
\noindent\textbf{Bloom Multifilter.}
The closest work related to ours is the Bloom Multifilter, devised by Xu et al. \cite{XuR16}, which extends standard Bloom Filter to check multiple elements on multiple sets at once. 

Let $\mathcal{S} = \{S_1, ..., S_n\}$, where each $S_i$ is a set of multiple elements. To check whether there is an $S \in \mathcal{S}$ which contains all the elements in a query $q$ we need to implement a Boolean function $f$ defined as:
\[
    f(q) =
    \begin{cases}
        1 & \exists S \in \mathcal{S} : q \subseteq S \\
        0 & \text{otherwise}
    \end{cases}
\]

Similarly to the Bloomier Filter, the Bloom Multifilter uses a bit matrix to represent multiple sets, and each set has an assigned ID. Whenever an element is added to such set, it is mapped to the rows having the indices equal to its hash neighborhood; the ID of the set, represented in binary, is then added to such rows using the bit-wise OR operation.

To check whether multiple elements belong to one of the sets, they are mapped to multiple rows according to their hash neighborhood, then the bit-wise AND operation is performed on such rows. If the result is a value greater than 0, then all those elements are probably in one of the sets.

Both standard Bloom Filter and Bloomier Filter\cite{CKR04} were meant to encode only one set, and the Bloom Multifilter~\cite{XuR16} can only answer whether it is true or false that one or multiple elements are in one of the represented sets. In this article, we have extended standard Bloom Filter not only to encode multiple sets and to efficiently check the membership of an element in all the sets but also to answer to which of the sets such element belongs to. Besides, a number of new approaches aim to model the input data, that a standard Bloom filter is going to present, with a pre-filter. These approaches use machine learning~\cite{Kraska:2018:CLI} or rely on standard Bloom Filter for pre-filtering~\cite{2018arXiv180301474M}. Our Bloom Test also can be used for pre-filtering data with multiple sets. 



\section{Conclusions} \label{conclusions}

In this article, we presented two statistical data structures which are able to answer not only the membership of the labels but also can answer to which sets they are associated with. At the same time, they are space-efficient and thus can be cached in RAM where the replication is less expensive in terms of storage. With randomly distributed labels amongst the sets, the variant of Bloom Matrix i.e., Sparse Bloom Matrix, is more space efficient at an expense of an reordering cost. With Zipf distributed labels amongst the sets, Bloom Vector is the most space efficient structure. Therefore, these structures are also statistically meaningful. We evaluated their performance for basic Bloom Filter operations. Finally, we introduced \emph{Bloom Test} to find whether the input sets together follow Uniform distribution or not. The test result can be used to determine which structure is suitable to achieve an FP rate of less than $10^{-2}$. 



\section*{Acknowledgements}
We would like to thank Dr. Antonio Cano, who worked on early Bloom Multifilter designs while visiting Helsinki.


\vspace{-1cm}
\begin{IEEEbiography}[{\includegraphics[width=1.0in,height=1.2in,clip, keepaspectratio]{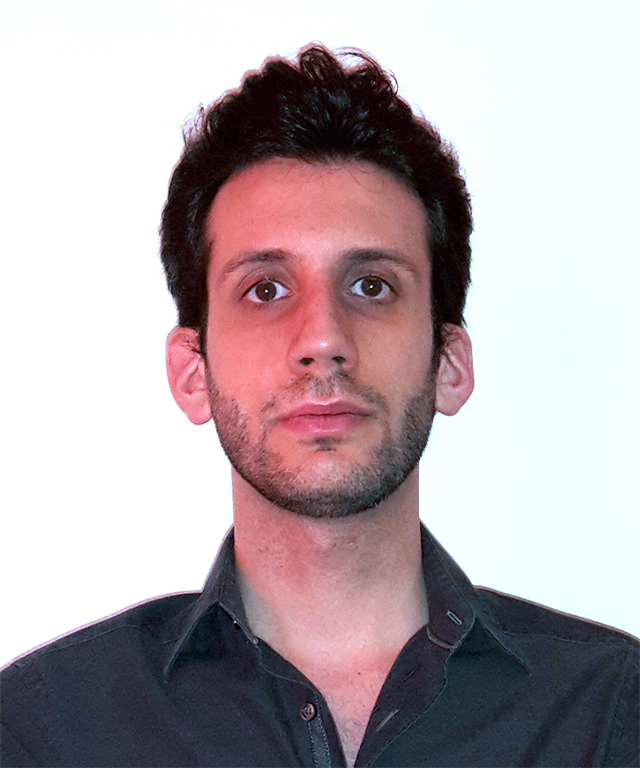}}]{Francesco Concas}
is a doctoral student in Computer Science at the University of Helsinki. His main research interests are algorithms and machine learning, and in particular probabilistic models.
\end{IEEEbiography}

\vspace{-1.2cm}
\begin{IEEEbiography}[{\includegraphics[width=1in,height=1.2in,clip,keepaspectratio]{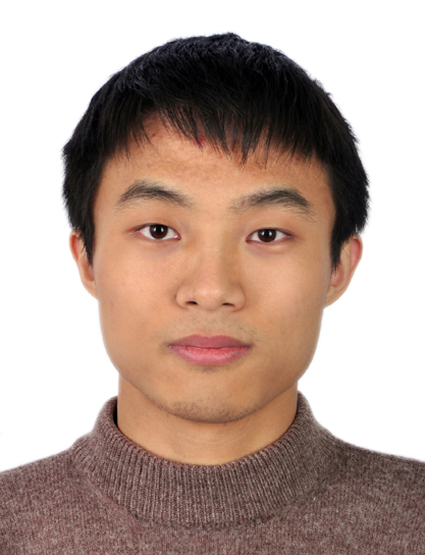}}]{Pengfei Xu}
is a doctoral student in the Department of Computer Science, University of Helsinki. His current research topic are string processing, DBMS query optimization, and Big Data algorithms.
\end{IEEEbiography}

\vspace{-1.2cm}
\begin{IEEEbiography}[{\includegraphics[width=1in,height=1.2in,clip,keepaspectratio]{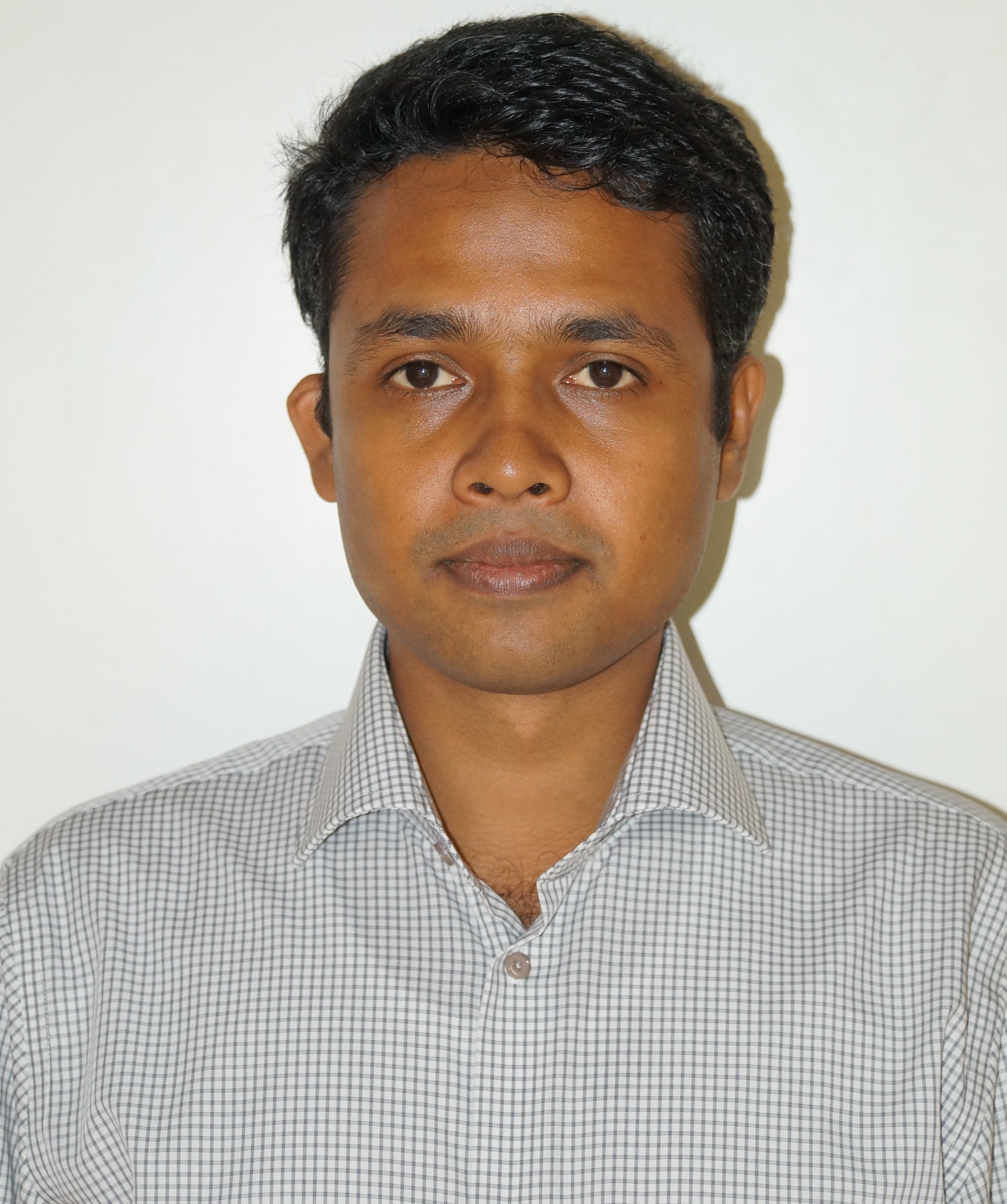}}]{Mohammad A. Hoque}
is a postdoctoral researcher at the  University of Helsinki. He obtained his M.Sc degree in Computer Science and Engineering in 2010, and Ph.D in 2013 from Aalto University. His research interests include energy efficient mobile computing, data analysis, distributed computing, and resource-aware scheduling.
\end{IEEEbiography}

\vspace{-1.2cm}
\begin{IEEEbiography}[{\includegraphics[width=1in,height=1.2in,clip,keepaspectratio]{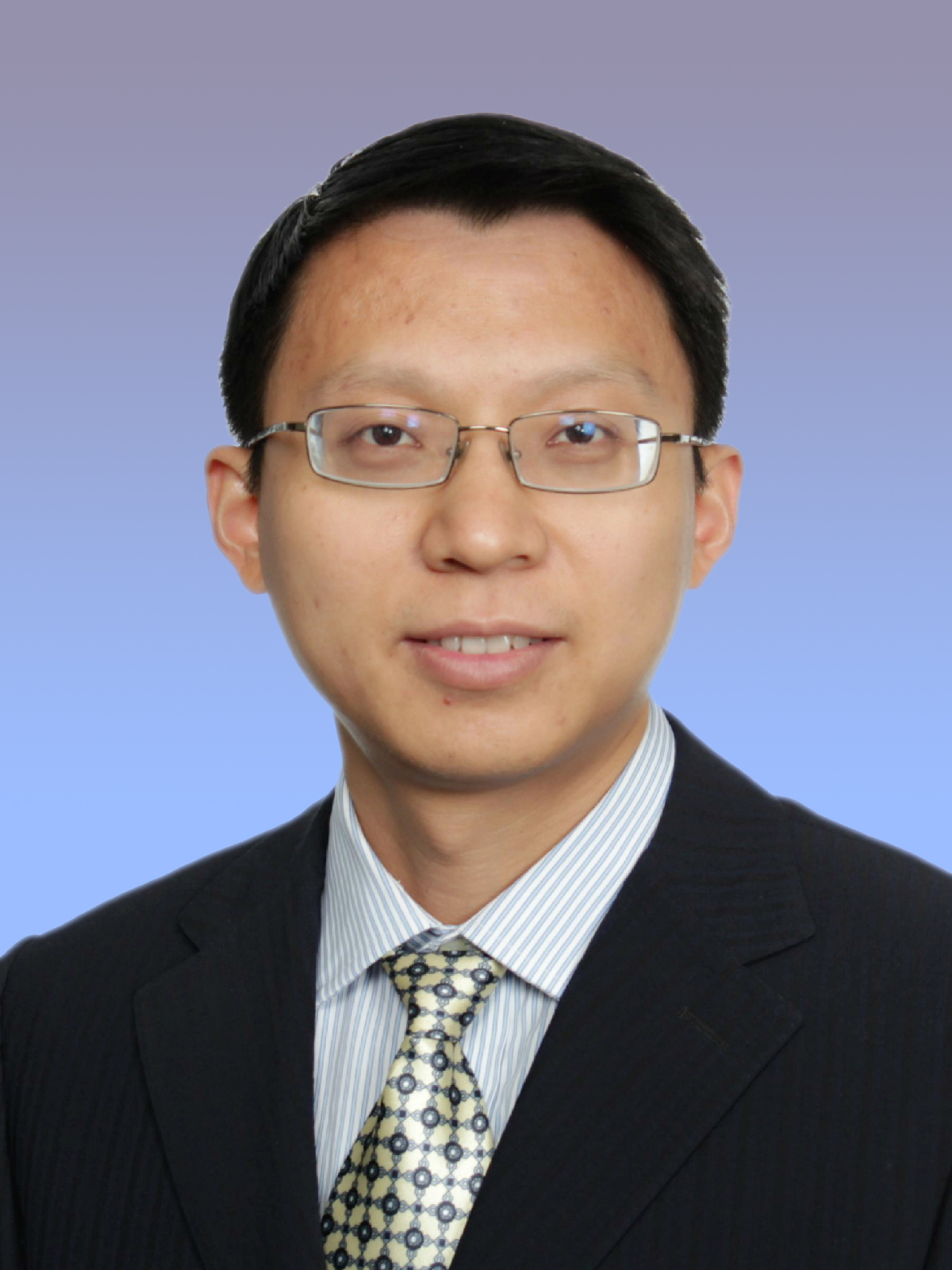}}]{Jiaheng Lu}
is an Associate Professor of the Department of Computer Science at the University of Helsinki, Finland.  His recent research interests include multi-model database management systems and job optimization for big data platform.
\end{IEEEbiography}

\vspace{-1.2cm}
\begin{IEEEbiography}[{\includegraphics[width=1in,height=1.2in,clip,keepaspectratio]{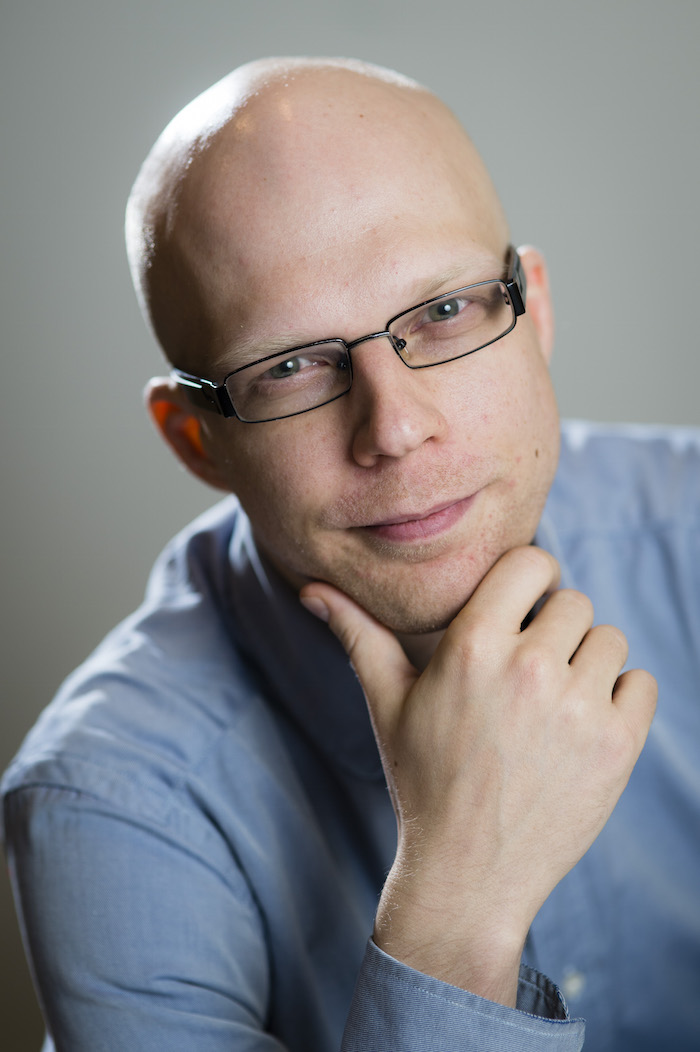}}]{Sasu Tarkoma}
(SMIEEE’12) is a Professor of Computer Science at the University of Helsinki, and Head of the Department of Computer Science. He has authored 4 textbooks and has published over 160 scientific articles.  His research interests are Internet technology, distributed systems, data analytics, and mobile and ubiquitous computing.  He has seven granted US Patents. His research has received several Best Paper awards and mentions, for example at IEEE PerCom, ACM CCR, and ACM OSR.
\end{IEEEbiography}

\end{document}